\documentclass{article}
\usepackage{setspace}
\usepackage[margin=3cm,nohead]{geometry}
\usepackage{appendix,graphicx}
\usepackage{amsfonts,amssymb,amsmath,mathenv}
\allowdisplaybreaks[3]

\newcommand{\nc}{\newcommand}
\def\foot{\footnote}
\def \bi{\bibitem}
\def \ci{\cite}
\nc{\sr}{\sqrt}
\nc{\fr}{\frac}
\nc{\ov}{\over}
\nc{\x}{\times}
\nc{\cosec}{\textrm{\,cosec\,}} 
\nc{\sech}{\textrm{\,sech\,}}
\nc{\cosech}{\textrm{\,cosech\,}}
\nc{\del}{\partial}
\nc{\dpl}{\partial_+}
\nc{\dm}{\partial_-}
\nc{\dpp}{\partial_+\partial_+}
\nc{\dmm}{\partial_-\partial_-}
\nc{\dpm}{\partial_+\partial_-}
\nc{\ra}{\rightarrow}
\nc{\lra}{\leftrightarrow}
\nc{\Ra}{\Rightarrow}
\nc{\til}{\tilde}
\nc{\R}{\mathbb{R}}
\nc{\Z}{\mathbb{Z}}
\nc{\C}{\mathbb{C}}
\nc{\al}{\alpha}
\nc{\bet}{\beta}
\nc{\ga}{\gamma}
\nc{\de}{\delta}
\nc{\h}{\eta}
\nc{\thet}{\theta_{\!_A}}
\nc{\ka}{\kappa}
\nc{\lam}{\lambda}
\nc{\si}{\sigma}
\nc{\Si}{\Sigma}
\nc{\ta}{\tau}
\nc{\ze}{\zeta}
\nc{\y}{\psi}
\nc{\om}{\omega}
\nc{\vt}{\vartheta}
\nc{\vp}{\varphi}
\nc{\be}{\begin{equation}}
\nc{\ee}{\end{equation}}
\nc{\bd}{\begin{displaymath}}
\nc{\ed}{\end{displaymath}}
\nc{\ba}{\begin{array}}
\nc{\ea}{\end{array}}
\nc{\la}{\label}

\nc{\nin}{\noindent}
\nc{\hs}{\hspace}
\nc{\vs}{\vspace}
\nc{\np}{\vspace{10pt} \\}
\nc{\mc}{\mathcal}
\nc{\mf}{\mathfrak}
\nc{\mbb}{\mathbb}
\nc{\trm}{\textrm}
\nc{\tbf}{\textbf}
\nc{\mbf}{\mathbf}
\def \Str {{\rm STr}}
\nc{\gi}{g^{-1}}
\nc{\finv}{f^{-1}}
\nc{\YL}{\Psi_{_L}}
\nc{\YR}{\Psi_{_R}}
\nc{\YLo}{\Psi_{_L0}}
\nc{\YRo}{\Psi_{_R0}}
\nc{\hm}{\h^\parallel}
\nc{\hh}{\h^\perp}
\nc{\T}{\mc{T}}
\newcommand{\commut}[2]{\left[ #1{,}\,#2 \right] }
\def \ha {\fr{1}{2}}
\def \la {\label} 
\def \bea {\begin{eqnarray}}
\def \eea {\end{eqnarray}}

\def \s {\sigma} 
 \def \k {\kappa}

\begin{document}

\overfullrule=0pt
\parskip=2pt
\parindent=12pt
\headheight=0in \headsep=0in \topmargin=0in \oddsidemargin=0in

\vspace{ -3cm}
\thispagestyle{empty}
\vspace{-1cm}

\rightline{Imperial-TP-2010-YI-01}

\rightline{    }

\begin{center}
\vspace{1cm}
{\Large\bf  
One-loop corrections to ${\rm AdS}_5\times {\rm S}^5$ superstring partition function via Pohlmeyer reduction

\vspace{1.2cm}

   }

\vspace{.2cm} {
Yukinori Iwashita\footnote{ yukinori.iwashita07@imperial.ac.uk }  }\\

\vskip 0.6cm

\textit{ 
 Theoretical Physics Group \\
 Blackett Laboratory, Imperial College\\
London SW7 2AZ, U.K. }

\end{center}

\onehalfspacing
\setcounter{footnote}{0}
\begin{abstract}

We discuss semiclassical expansions around a class of classical string configurations lying in ${\rm AdS}_3\times {\rm S}^1$ using the Pohlmeyer-reduced from of the 
${\rm AdS}_5\times {\rm S}^5$ superstring theory. 
The Pohlmeyer reduction of the ${\rm AdS}_5\times {\rm S}^5$ superstring theory is a gauged Wess-Zumino-Witten model with an integrable potential and two-dimensional fermionic fields.  
It was recently conjectured that the quantum string partition function is equal to the quantum reduced theory partition function. 
Continuing the previous paper (arXiv:0906.3800) where arbitrary solutions in ${\rm AdS}_2\times {\rm S}^2$ and homogeneous solutions were considered, 
we provide explicit demonstration of this conjecture at the one-loop level for several string solutions in ${\rm AdS}_3\times {\rm S}^1$ embedded into ${\rm AdS}_5\times {\rm S}^5$.  
Quadratic fluctuations derived in the reduced theory for inhomogeneous strings are equivalent to respective fluctuations found from the Nambu action in the original string theory.   
We also show the equivalence of fluctuation frequencies for homogeneous strings with both the orbital momentum and the winding on a big circle of ${\rm S}^5$.

\end{abstract}

\newpage
\tableofcontents

\setcounter{footnote}{0}

\renewcommand{\theequation}{1.\arabic{equation}}
\setcounter{equation}{0}
 \newpage

\section{Introduction}

With a motivation to understand quantum ${\rm AdS}_5 \times {\rm S}^5$ superstring theory, 
here we discuss semiclassical quantization of the Pohlmeyer-reduced form of a coset sigma model on $\frac{PSU(2,2|4)}{Sp(2,2)\times Sp(4)}$.

The Pohlmeyer reduction was proposed to show that the equation of motion of the chiral ${\rm S}^2$ sigma model is reduced to the sin-Gordon equation \cite{pohl}. 
The application of this technique to conformal-gauge bosonic string theory includes uncovering the integrability of classical string motion in de Sitter spacetime \cite{devega} and finding classical string configurations localized in subspace of ${\rm AdS}_5 \times {\rm S}^5$ \cite{dor,okasuz,hosv,klomac,jevjin1,jevjin2,mira} 
as the integrable property gives us powerful methods to generate soliton solutions in the reduced theory and the string solution can be constructed from a solution in the reduced theory. 
In particular string theory in ${\rm R}\times {\rm S}^3$ and string theory in ${\rm AdS}_3\times {\rm S}^1$ are reduced to the complex sin-Gordon model and the complex sinh-Gordon model, respectively, and the latter is the case which we consider in this paper. 
More recently the Pohlmeyer reduction of ${\rm AdS}_3$ string theory was used in evaluating the minimal area of an open string surface ending on null Wilson loop, which is related to the strong-coupling limit of ${\cal N}=4$ super Yang-Mills theory \cite{almal1, misu1,misu2}. 
For this sector it is known that the reduced form is the generalized sinh-Gordon model. 
Although one can easily switch off the ${\rm S}^1$ sector of most of the classical string solutions stretching in ${\rm AdS}_3\times {\rm S}^1$, 
the relation between the complex sinh-Gordon model and the generalized sinh-Gordon model is understood only at the level of equation of motion.\footnote{
In the reduced theory the case of no motion or no stretching in ${\rm S}^1$ corresponds to taking $\mu \to 0$ where $\mu $ is the mass scale of the reduced theory. 
It is pointed out in \cite{gt2} that one may be able to take this limit at the level of the gWZW Lagrangian of the full reduction, but so far, the $\mu \to 0$ limit at the level of classical Lagrangian is not well understood. 
As we will see later, this limit is well defined in the fluctuation Lagrangian.   
} 
Open string solutions stretching in ${\rm AdS}_n$ with $n>3$ or ${\rm AdS}_3\times {\rm S}^3$ are discussed in \cite{almal2, misu1,misu2,misu3}.

The Pohlmeyer-type reduction of the full ${\rm AdS}_5 \times {\rm S}^5$ superstring sigma model was proposed in \cite{gt1,mikhshaf,gt2}.
Type IIB Green-Schwarz superstring action in ${\rm AdS}_5 \times {\rm S}^5$ can be formulated as a sigma model action on the coset superspace $F/G=PSU(2,2|4)/(Sp(2,2)\times Sp(4))$ \cite{mt}. 
In the construction of the reduced theory, new variables are algebraically related to supercoset current components and the Virasoro conditions are automatically solved, and the local fermionic $\kappa$ symmetry is fixed.
The resulting system is expressed as gauged Wess-Zumino-Witten (gWZW) model associated with $G/H=(Sp(2,2)\times Sp(4))/[SU(2)]^4$ and deformed with an integrable potential and two dimensional fermionic fields. 
The reduced Lagrangian exhibits the two-dimensional Lorentz invariance, and after integrating out the gauge fields, the reduced theory involves $8$ bosonic degrees of freedom and $16$ fermionic degrees of freedom, i.e., involves only physical degrees of freedom. 
The deformed gWZW model possesses the mass scale $\mu$ which is to be extracted from the stress tensor in the AdS sector of the original string theory,
\be
T_{\pm \pm} ^{\rm AdS}=-\mu^2_\pm \,.
\ee
Because the closed string theory is defined on a cylinder rather than a plane, One is not allowed to use the 2d Lorentz symmetry in order to set $\mu_+=\mu_-$. 
Even for the $\mu_+\ne \mu_-$ case, however, we can use a single $\mu$ defined by $\mu =\sqrt{\mu_+\mu_-}$.\footnote{
Equivalently we obtain $\mu_ \pm$ from the other sector, $T^{\rm S}_{\pm \pm}=\mu_\pm^2 $, since they are connected by the Virasoro constraints, $T_{\pm \pm} ^{\rm AdS}+T^{\rm S}_{\pm \pm}=0$. 
As far as the sector of ${\rm AdS}_3\times {\rm S}^1$ is concerned, it is more convenient to calculate $\mu_\pm$ in the ${\rm S}^1$ sector due to its simple structure. 
To be precise, $\mu_{\pm}$ are not necessarily constant because the conservation of the stress tensor implies $\partial _\pm T_{\mp \mp}=0$, i.e., $T_{\pm \pm}^{\rm AdS}=-[f_{\pm}(\sigma^{\pm})]^2$. So we can not set $T_{\pm \pm} ^{\rm AdS}$ to be constant without using the 2d conformal invariance. 
All examples of string solutions in this paper are the case of constant $\mu _\pm$, and we do not encounter this problem.   

One possibility to find a solution with nonconstant $\mu_\pm$ is to employ the ansatz $\partial_-X=0$ and $\partial_-Y=0$ with $\partial_+X\ne 0$ and $\partial_+Y\ne 0$. 
With this choice the stress tensor is $T_{++} ^{\rm AdS}=T_{++} ^{\rm S}=0$ and $T_{--} ^{\rm AdS}=-T_{--} ^{\rm S}=-[f(\sigma ^-)]^2$, i.e., we have $\mu_+=0$ and $\mu_-=f(\sigma ^-)$. 
}

Since conformal invariance on the string worldsheet plays a crucial role in the Pohlmeyer reduction, the classical equivalence of the original ${\rm AdS}_5 \times {\rm S}^5$ superstring theory and the deformed gWZW model can be extended to the quantum level only if the two theories are UV finite. 
In fact the UV finiteness of the ${\rm AdS}_5 \times {\rm S}^5$ superstring sigma model was proved up to two-loop order \cite{rtt,rotecusp,rt2}, and the 
reduced theory is also UV finite at the same order \cite{rtfin}. 
In \cite{hit} we conjectured that the quantum partition functions of the original string theory and the reduced theory are equal, 
\be
 \mathcal{Z}_{\bf string \ theory}^{^{(q)}} = \mathcal{Z}_{\bf reduced\ theory} ^{^{(q)}}  \la{eqpa} \, ,
\ee 
and demonstrated its validity at the one-loop level for any strings localized in the ${\rm AdS}_2 \times {\rm S}^2$ subspace, the homogeneous folded string in ${\rm AdS}_3 \times {\rm S}^1$, and the homogeneous circular two-spin string in ${\rm R}_1 \times {\rm S}^3$.
In this class the classical solutions always have $\mu _+= \mu_-$, then we immediately have $\mu=\mu _\pm $ in the reduced theory.

The purpose of this paper is to verify the conjecture \eqref{eqpa} for more nontrivial string solutions in ${\rm AdS}_3 \times {\rm S}^1$ by deriving their fluctuation Lagrangians in the reduced theory and comparing them to the string theory result. 
Also we shall show the equivalence of fluctuation frequencies for homogeneous string solutions 
and the smoothness of the $\mu \to 0$ limit at the level of the fluctuation Lagrangian for all cases. 
Because the reduced theory has a simpler quantum structure than the original string theory, 
semiclassical computation in the reduced theory is turned out to be much easier for both bosons and fermions. 
Among many classical string configurations stretching in the ${\rm AdS}_3 \times {\rm S}^1$ subspace we shall consider the folded string, the circular string and the spiky string.

This paper is organized as follows.

We shall begin with a brief review on the Pohlmeyer reduction and its perturbation theory in section 2. 
For the bosonic string theory in ${\rm AdS}_3 \times {\rm S}^1$ one can explicitly write the relation between the embedding coordinates in the original string theory and two fields of the reduced theory. There we will introduce two reduced models called the coth model and the tanh model, which are connected by the ``T-duality'' transformation.  
We shall show that these two models are embedded in the reduced model of the ${\rm AdS}_5 \times {\rm S}^5$ superstring theory and  
are related by the $H\times H $ gauge transformation rather than the $H$ gauge transformation in the full reduction.\footnote{
Here we have $H=[SU(2)]^4$.
Because $H\times H $ is the symmetry of the equations of motion for the reduced theory, 
the coth model and the tanh model are not connected at the level of the gWZW Lagrangian. 
This is the same as the situation in the complex sinh-Gordon model. 
The T-duality transformation is defined at the level of the sinh-Gordon equations.
}

In section 3 we shall discuss the $(S,J)$ folded string. 
A folded string in pure ${\rm AdS}_3$ was first studied as the simplest string state whose classical energy grows
logarithmically with the spacetime spin in ${\rm AdS}_3$ \cite{gkp}, and soon after, this solution was extended to the $(S,J)$ folded string solution in ${\rm AdS}_3 \times {\rm S}^1$ where the ${\rm S}^1$ sector is the pointlike string moving along a big circle of ${\rm S}^5$, i.e., the BMN state \cite{ft1}. 
Its quadratic fluctuations were found from the Nambu action in the static gauge and in the Polyakov action in the conformal gauge \cite{ft1, ftt}, 
and the equivalence of these two approaches was shown at the one-loop level in \cite{bdfpt}. 
Here one may ask which type of the fluctuation Lagrangian arises in the reduced theory. 
We shall show the fluctuation Lagrangian derived from the Nambu action is related to that of the coth model.

In section 4 we shall study the homogeneous $(S,J)$ circular string solution which has both the angular momentum and the winding on a big circle of ${\rm S}^5$, and gives us the first example of $\mu_+ \ne \mu_-$. 
In \cite{art,ptt,gv} semiclassical expansions around the circular string were worked out, and in particular, its fermionic fluctuations are evaluated by carrying out the worldsheet computation in \cite{ptt} and by employing the algebraic curve method in \cite{gv}.\footnote{ 
The sum over fermionic frequencies should be taken over integer mode number in the former approach or half-integer mode number in the latter approach. 
It was shown in \cite{mikhcir} that this discrepancy is resolved by carefully considering the spin bundle over the ${\rm AdS}_5 \times {\rm S}^5$ spacetime in the worldsheet computation. 
}
We shall show the fluctuation Lagrangian of the reduced theory agrees with \cite{ptt}. 
Because the circular string is homogeneous, we can evaluate characteristic frequencies of its quadratic fluctuations also in the reduced theory. 
In the case of the $(S,J)$ circular string the total sum of the bosonic and fermionic frequencies in the reduced theory agrees with the string theory result although some of the individual frequencies appear to disagree.

A spiky string solution in ${\rm AdS}_3$ was first found in \cite{krusp} as a generalization of the folded string solution, and extended to a solution stretching in ${\rm AdS}_3 \x {\rm S}^1$ in \cite{iktt}. 
In section 5 we shall discuss the $(S,J)$ spiky string solution and derive its fluctuation Lagrangian of the reduced theory. 
Our result on the semiclassical expansions for the $(S,J)$ spiky string is expected to agree with the one in the original string theory. 
As the $(S,J)$ spiky string solution is also the case of $\mu_+ \ne \mu_-$, we consider that the discrepancy could happen for the individual fluctuations, but the total sum the fluctuations should be the same as in the string theory.
We shall also discuss the fluctuation Lagrangian in the limiting cases of the $(S,J)$ spiky string; spiky string without motion or stretching in ${\rm S}^5$, and its folded string limit.

In the large spin limit, spikes of the spiky string approach the conformal boundary of ${\rm AdS}_3$, and the solution becomes locally equivalent to the scaling limit of the folded string. 
The existence of this homogeneous limit is shown in \cite{iktt} where the expression for its string energy is similar to that for the homogeneous $(S,J)$ folded string. 
A generalization of the $(S,J)$ folded string solution leads to the explicit construction of another limiting solution of the $(S,J)$ spiky string; the new folded string has both the orbital momentum and the winding in the ${\rm S}^1$ sector \cite{grrtv}. 
This solution again has the scaling limit where the solution becomes homogeneous, which is the case we shall study in section 6.  
As happened in the circular string case, some individual characteristic frequencies in the reduced theory disagree with the result in \cite{grrtv}, but the sum of the frequencies is the same as in the string theory.

Some concluding remarks are made in section 7.

The $\mathfrak{psu}(2,2|4)$ superalgebra is summarized in appendix \ref{appsupcos}, which will be used when we introduce component fields of the fluctuations in the reduced theory.    
In appendix \ref{apppuls} we shall study fluctuations around the pulsating string solution \cite{krutse,mincir} in ${\rm R} \times {\rm S}^2$ in the reduced theory. 
Since the reduced theory for the ${\rm R} \times {\rm S}^2$ string theory is the sin-Gordon model for a single field, 
one can check that the fluctuation Lagrangian of the reduced theory is exactly the same as that found from the Nambu action in the original string theory.

\renewcommand{\theequation}{2.\arabic{equation}}
 \setcounter{equation}{0}

\section{Review of Pohlmeyer reduction}

In this section we shall describe the perturbation in the reduced Lagrangian.  
In \ref{tancotwzw} we shall first review the Pohlmeyer reduction of bosonic string theory in ${\rm AdS}_3 \times {\rm S}^1$, and particularly, focus on two reduced  models called the coth model and the tanh model of the complex sinh-Gordon theory, which are related by the ``T-duality'' transformation as discussed in detail in \cite{gt1} for the case of complex sin-Gordon model.  
Then we shall derived the Lagrangian for quadratic fluctuations in the both models. 
In \ref{pohlfull} we shall review the perturbation in the deformed 
gWZW model studied in \cite{hit}, and discuss the embedding of the coth model and the tanh model into the full reduction.

\subsection{Pohlmeyer reduction of bosonic string theory in ${\rm AdS}_3 \times {\rm S}^1$ \label{tancotwzw}}

Our starting point is the worldsheet Lagrangian for a bosonic string propagating in ${\rm AdS}_3 \times {\rm S}^1$ spacetime, 
\begin{equation}
\mathcal{L}=
\mathcal{L}_{\rm AdS} +
\mathcal{L}_{\rm S}\,,
\end{equation} 
with
\begin{equation} \label{strac}
\begin{array}{c}
\mathcal{L}_{\rm AdS} = \partial_+ Y^P \partial _-Y_P- \Lambda \left( Y^PY_P+1 \right) \,, \\
\mathcal{L}_{\rm S}=\partial_+ X^M \partial _-X_M-\tilde{\Lambda} \left( X^MX_M-1 \right) \,,
\end{array}
\end{equation} 
where $\partial_\pm=\partial_{\tau} \pm \partial_{\sigma}$ 
and the contraction is defined by using $\eta={\rm diag}(-1,1,1,-1)$ for $P,Q,\dots=0,1,2,3$ indices (${\rm AdS}_3$ sector)
and $\delta={\rm diag}(1,1)$ for $M,N,\dots=1,2$ indices (${\rm S}^1$ sector). 

Reflecting the fact that the ${\rm S}^1$ sector of a string solution in ${\rm AdS}_3 \times {\rm S}^1$ is always homogeneous, the AdS part of the stress tensor satisfies $T_{\pm \pm}^{\rm AdS}=-\mu_{\pm}^2$ with constant $\mu_{\pm}$.
Because the worldsheet of a closed string is a cylinder, it is not necessarily allowed to set 
$\mu_+ = \mu_- \equiv \mu$.\footnote{We have $\mu_+=\mu-$ in several cases, e.g., when the ${\rm S}^1$ sector is the BMN vacuum.} 
Instead we introduce the mass scale in the reduced theory by $\mu =\sqrt{\mu_+\mu_-}$. This prescription will be used in the case of classical string with both the orbital momentum and the winding in the ${\rm S}^1$ sector (see sections \ref{sjcs}, \ref{sjspik}, \ref{mwfold}). 

In the case of the ${\rm AdS}_3 \times {\rm S}^1$ bosonic string theory, we can explicitly write the relation between the embedding coordinates and scalar fields of the reduced theory.
Let us first look at the construction of the coth model. 
Introduce a set of $O(2,2)$ vectors by $Y_P$, $\partial_+Y_Q$, $\partial_-Y_R$ and $K_P \equiv \epsilon_{QRSP}Y^Q\partial_+Y^R\partial_-Y^S$, and 
define $\phi$ and $\theta_{\!_A}$ by 
\begin{equation}\label{redcoth}
\begin{array}{c}
\partial_+ Y^P \partial _-Y_P=-\mu ^2 \cosh 2\phi_{\!_A} \,, \\
K_P\,\partial_\pm^2Y^P= 4\mu ^3\cosh ^2\!\phi_{\!_A} \, \partial_\pm \chi_{\!_A} \,,
\end{array}
\end{equation} 
then these $\phi_{\!_A}$ and $\chi_{\!_A}$ satisfy the complex sinh-Gordon equations, 
\be \ba{c} 
	\partial_+ \partial_- \phi_{\!_A}+\frac{\cosh \phi_{\!_A}}{\sinh^3\phi_{\!_A}}\partial_+\chi_{\!_A} \partial_-\chi_{\!_A} +\frac{1}{2}\mu^2 \sinh2\phi_{\!_A}=0 \,, \\
	\partial_+ (\coth^2\phi_{\!_A} \,\partial_-\chi_{\!_A} )+\partial_- (\coth^2\phi_{\!_A} \,\partial_+\chi_{\!_A} )=0\,,
\ea \ee
which follow from the Lagrangian of the coth model, 
\be \label{comschc}
	\mathcal{L}_{\rm coth}=\partial_+\phi_{\!_A} \partial_- \phi_{\!_A}+\coth^2 \! \phi_{\!_A} \, \partial_+\chi_{\!_A} \partial_-\chi_{\!_A} -\frac{1}{2}\mu^2 \cosh2\phi_{\!_A} \,. \\
\ee

As we will see in the following sections, the perturbation in the complex sinh-Gordon model describes a part of the fluctuations in the deformed gWZW model. 
Moreover, it can be a useful tool for evaluating a complicated subsector of the physical fluctuations in the deformed gWZW model. 
By the perturbation, $\phi_{\!_A} \to \phi_{\!_A} +\delta \phi_{\!_A}$ and $\chi_{\!_A} \to \chi_{\!_A} +\delta \chi_{\!_A}$, in \eqref{comschc} we obtain the quadratic fluctuations, 
\be \label{quadcoth} \ba{c}
	\mathcal{L}_{\rm coth\, (2)}=\partial_+\delta \phi_{\!_A} \partial_- \delta \phi_{\!_A} 
	+\left( \frac{3+2 \sinh^2 \! \phi_{\!_A} }{\sinh^4 \! \phi_{\!_A} } \partial_+\chi_{\!_A} \partial_-\chi_{\!_A} -\mu^2 \cosh2\phi_{\!_A} \right)(\delta \phi_{\!_A})^2 \\
	\hspace{84pt}+\coth^2 \! \phi_{\!_A} \, \partial_+\delta\chi_{\!_A} \partial_-\delta\chi_{\!_A}  -\frac{2 \cosh \! \phi_{\!_A} }{\sinh^3 \! \phi_{\!_A} } (\partial_+\chi_{\!_A} \partial_-\delta\chi_{\!_A} +\partial_+\delta \chi_{\!_A} \partial_-\chi_{\!_A} )\delta\phi_{\!_A}
	 \,. \\
\ea \ee

Another model of the complex sinh-Gordon theory is the tanh model obtained by replacing \eqref{redcoth} by 
\begin{equation}\label{red}
\begin{array}{c}
\partial_+ Y^P \partial _-Y_P=-\mu ^2 \cosh 2\phi_{\!_A} \,, \\
K_P\,\partial_\pm^2Y^P=\mp 4\mu ^3\sinh ^2\!\phi_{\!_A} \, \partial_\pm \theta_{\!_A} \,.
\end{array}
\end{equation} 
The resulting equations describe the tanh model, 
\be \ba{c} 
	\partial_+ \partial_- \phi_{\!_A}-\frac{\text{sinh}\phi_{\!_A}}{\text{cosh}^3\phi_{\!_A}}\partial_+\theta_{\!_A} \partial_-\theta_{\!_A} +\frac{1}{2}\mu^2 \text{sinh}2\phi_{\!_A}=0 \,, \\
	\partial_+ (\text{tanh}^2\phi_{\!_A} \,\partial_-\theta_{\!_A} )+\partial_- (\text{tanh}^2\phi_{\!_A} \,\partial_+\theta_{\!_A} )=0\,,
\ea \ee
whose Lagrangian is 
\be \label{comsch}
	\mathcal{L}_{\rm tanh}=\partial_+\phi_{\!_A} \partial_- \phi_{\!_A}+\tanh^2 \! \phi_{\!_A} \, \partial_+\theta_{\!_A} \partial_-\theta_{\!_A} -\frac{1}{2}\mu^2 \cosh2\phi_{\!_A} \,. \\
\ee
As pointed out in \cite{gt1} for the complex sin-Gordon model, these two models are related at the level of equations of motion by the ``T-duality'' transformation,  
\be \label{tdual}
	\partial_{\pm}\chi_{\!_A} = \mp \tanh^2 \! \phi_{\!_A} \, \partial_{\pm}\theta_{\!_A} \,. 
\ee
Because this transformation is non-local, a classical solution in the reduced theory might take a complicated form in one model even if it is simple in the other model. 

It is also useful to discussing the perturbation in the tan model. 
By $\phi_{\!_A} \to \phi_{\!_A} +\delta \phi_{\!_A}$ and $\theta_{\!_A} \to \theta_{\!_A} +\delta \theta_{\!_A}$ in \eqref{comsch}, we obtain the Lagrangian for the quadratic fluctuations, 
\be \label{quadtanh} \ba{c}
	\mathcal{L}_{\rm tanh\, (2)}=\partial_+\delta \phi_{\!_A} \partial_- \delta \phi_{\!_A} 
	+\left( \frac{3-2\cosh^2 \! \phi_{\!_A} }{\cosh^4 \! \phi_{\!_A} } \partial_+\theta_{\!_A} \partial_-\theta_{\!_A} -\mu^2 \cosh2\phi_{\!_A} \right)(\delta \phi_{\!_A})^2 \\
	\hspace{78pt}+\tanh^2 \! \phi_{\!_A} \, \partial_+\delta\theta_{\!_A} \partial_-\delta\theta_{\!_A}  +\frac{2 \sinh \! \phi_{\!_A} }{\cosh^3 \! \phi_{\!_A} } (\partial_+\theta_{\!_A} \partial_-\delta\theta_{\!_A} +\partial_+\delta \theta_{\!_A} \partial_-\theta_{\!_A} )\delta\phi_{\!_A}
	 \,. \\
\ea \ee
Note that the the T-duality transformation works even at the semiclassical level, that is, \eqref{quadtanh} is T-dual to \eqref{quadcoth},
which we will use in section \ref{secfold}. 

On general ground there is no reason to expect that the complex sinh-Gordon theory is equivalent to the ${\rm AdS}_3 \times {\rm S}^1$ bosonic string theory at the quantum level. 
As far as the one-loop corrections are concerned, on may think this expectation might be true because the first order corrections directly follow from the equations of motion. 
However, this is not correct for the case of $\mu_+\ne \mu_-$, and it is necessary to take into account all the bosonic and fermionic fluctuations in order to obtain a correct set of physical fluctuations, which implies that the full reduction of the ${\rm AdS}_5 \times {\rm S}^5$ superstring is essential to extend the classical equivalence between string theory and its Pohlmeyer-reduced form into the semiclassical level.

\subsection{Pohlmeyer reduction of ${\rm AdS}_5 \times {\rm S}^5$ superstring\label{pohlfull}}

First we shall start with a brief summary of the construction of the reduced theory for the ${\rm AdS}_5 \times {\rm S}^5$ superstring theory. 
See \cite{gt1, gt2, hit} for its detail. 

The Lagrangian for the superstring in ${\rm AdS}_5 \times {\rm S}^5$ can be written in terms of the $\frac{F}{G}$ supercoset current components, 
where $F=PSU(2,2|4)$ and $G=Sp(2,2)\times Sp(4)$ (the $\mathfrak{psu}(2,2|4)$ superalgebra is reviewed in appendix \ref{appsupcos}). 
The variation of the Lagrangian yields a set of equations of motion, supplemented by the Virasoro constraints and the Maurer-Cartan equation. 
The Pohlmeyer reduction includes the processes of introducing new fields such that the Virasoro constraints are automatically solved and of fixing the $\kappa$ symmetry for the fermionic currents. 
The resulting system is governed by a group element $g\in G$, gauge fields $A_{\pm}$ taking value on the algebra $\mf{h}$ of the group $H=[SU(2)]^4$, and $16$ two-dimensional fermions $\YR \in \mf{f}_1^{\parallel}$, $\YL \in \mf{f}_3^{\parallel}$, which satisfy the following equations,  
\be\ba{l}
\la{redeom}
\dm\left(\gi\dpl g+\gi A_+ g\right)-\dpl A_- +\left[A_-,\gi \dpl g + \gi A_+ g\right] \\ 
 \hspace{120pt} =-\mu^2\left[\gi T g, T\right]- \mu\left[\gi\YL g, \YR\right]\,,
\\D_-\YR=\mu\left[T,\gi \YL g\right]\,,\hspace{30pt}D_+\YL=\mu\left[T,g \YR \gi\right]\,,
\ea\ee
where $D_\pm=\del_\pm+\left[A_\pm,\right]$. $T$ is a constant matrix chosen to be 
\be 
T=\fr{i}{2}\trm{ diag}\left(1,\, 1,\, -1,\, -1,\, 1,\, 1,\, -1,\, -1\right)\,.
\ee
The equations \eqref{redeom} have  $H\x H$  gauge symmetry,
\be\ba{c}\la{hxhgauge}
g\ra h^{-1}g\bar{h}\,,\hs{20pt}A_+\ra h^{-1}A_+h+h^{-1}\dpl h,\,\hs{20pt}A_-\ra \bar{h}^{-1}A_-\bar{h}+\bar{h}^{-1}\dm \bar{h}
\\\YR\ra\bar{h}^{-1}\YR \bar{h}\,,\hs{20pt}\YL\ra h^{-1}\YL h\,, \hs{20pt}\ h,\bar{h} \in H \,. \ea\ee

In order to write down a Lagrangian for the equations of motion \eqref{redeom} we should partially
 fix the $H\x H$ gauge symmetry to a  $H$ gauge symmetry as in \ci{hit}, 
\be
\la{redgaugef}\ba{l}
\tau\left(A_+\right)=\left(\gi\dpl g+\gi A_+ g-\fr{1}{2}\left[\left[T,\YR\right],\YR\right]\right)_\mf{h}\,,
\\ \tau^{-1}\left(A_-\right)=\left(-\dm g \gi+g A_- \gi -\fr{1}{2}\left[\left[T,\YL\right],\YL\right]\right)_\mf{h}\,.
\ea\ee
Here $\tau$ is a supertrace-preserving\foot{$\trm{STr}\left(\tau\left(u_1\right)
\tau\left(u_2\right)\right)=\trm{STr}\left(u_1u_2\right)$, \ $u_{1,2}\in\mf{h}$.} 
automorphism of the algebra $\mf{h}$. 
This partial gauge fixing reduces the $H \x H$ gauge symmetry to the following asymmetric $H$ gauge symmetry,
\be\ba{l}\la{hgauge}
g\ra h^{-1}g\hat{\tau}\left(h\right)\,,\\ 
A_+\ra h^{-1}A_+h+h^{-1}\dpl h,\,\hs{10pt}A_-\ra 
\hat{\tau}\left(h\right)^{-1}A_-\hat\tau\left(h\right)+\hat{\tau}\left(h\right)^{-1}\dm \hat\tau\left(h\right) \,,
\\ \YR\ra \hat{\tau}\left(h\right)^{-1}\YR \hat\tau\left(h\right)\,,\hs{20pt}\YL\ra h^{-1}\YL h\,, \hs{20pt}\ h \in H \,, \ea\ee
where $\hat{\tau}$ is a lift of $\tau$ from $\mf{h}$ to $H$. 
The equations of motion, \eqref{redeom}, and the gauge field equations, \eqref{redgaugef},
 follow from the Lagrangian,
\begin{equation}\label{Ltot}
	\begin{split}
		\qquad \qquad L_{\rm dWZW}&=L_{\rm gWZW}+\mu^2\,\Str(g^{-1}Tg T )\\
		&~~~+{\textstyle \frac{1}{2}}\mathrm{STr}\left(\Psi_{_L}\commut{T}{D_+\Psi_{_L}}+
		\Psi_{_R}\commut{T}{D_-\Psi_{_R}}\right) +\ \mu\,
		\mathrm{STr}\left(
		g^{-1}\Psi_{_L}g\Psi_{_R}\right)\,,
	\end{split}
\end{equation}
where $L_{\rm gWZW}$ is the Lagrangian  of  the asymmetrically  gauged  $G/H$ WZW model,
\begin{equation} \label{gauwzw}
	\begin{split}
		I_{\rm gWZW}  &= \int\frac{d^2\sigma}{4\pi}{\rm STr}(g^{-1}\partial_+g g^{-1}\partial_-g) - \int\frac{d^3\sigma}{12\pi}
		{\rm STr}(g^{-1}dg g^{-1}dg g^{-1}dg)\\
		&~~~+~\int \fr{d^2 \sigma}{2\pi} \Str \left(  A_+\,
		\partial_- g g^{-1} -
		A_- \,g^{-1}\partial_+ g-g^{-1} A_+ g  A_-  + \tau\left(A_+\right) A_- \right) \,.
	\end{split}
\end{equation}
This Lagrangian is invariant under the gauge transformations \eqref{hgauge}.

 Let us now consider the fluctuations around a classical solution, $g_0$, $A_{0\pm}$, $\Psi_{_R0}$, $\Psi_{_L0}$, as follows 
\be \ba{l}
	g=g_0 e^\h = g_0(1+\h +\ha \h^2 +\mc{O}(\h^3))  \,,
	\\ A_+=A_{+0}+\de A_+ \,,~~~~~~ A_-=A_{-0}+\de A_- \,,
	\\ \Psi_{_R}=\Psi_{_R0}+\de \Psi_{_R} \,, ~~~~~~ \Psi_{_L}=\Psi_{_L0}+\de \Psi_{_L}\,.
\ea\ee
Hereafter we will consider classical solutions with vanishing fermions, i.e., $\Psi_{_R0}=\Psi_{_L0}=0$. 
Under this perturbation the quadratic fluctuations of the Lagrangian for the deformed gWZW model \eqref{Ltot} are described by 
\begin{equation} \label{quafluc}
	\ba{l}
	 \mathcal{L}_{_{\rm dWZW (2)}}={\rm STr} \bigg[ \ha \dpl \h \dm \h + \ha \left( \h \dm \h -\dm \h \h \right) g_0^{-1}\! \dpl g_0 +\de A_+ g_0 \dm \h \gi _0  
	- \ha A_{0+}g_0 \dm \h \h \gi _0 \hspace{50pt} \ \\ 
	  \hspace{50pt} + \ha A_{0+} g_0 \h \dm \h \gi _0 + \de A_- \h \gi _0 \! \dpl g_0  -\de A_-\gi_0  \! \dpl g_0 \h -\de A_- \dpl \h   + A_{0-} \h \gi _0 \dpl g_0 \h  \bigg. \\  
	\hspace{50pt} + \ha A_{0-}\h \dpl \h -\ha A_{0-}\h ^2 \gi _0 \! \dpl g_0   -\ha A_{0-}\gi_0  \dpl g_0 \h ^2 -\ha A_{0-}\dpl \h \h  
	+\de A_+ \de A_- \bigg. \\
   \hspace{50pt} -\ha \h ^2 \gi _0 A_{0+} gA_{0-} -\ha \gi _0 A_{0+} g_0 \h ^2 A_{0-} +
   \h \gi _0 \de A_+ g_0 A_{0-}   + \h \gi _0 A_{0+}g_0 \h A_{0-} \bigg. \\ 
	\hspace{50pt} 
    +\h \gi _0 A_{0+}g_0 \de A_-     - \gi _0  \de A_+ g_0 \h A_{0-} -\gi _0\de A_+
    g_0\de A_-  -\gi _0A_{0+} g_0 \h \de A_- \bigg. \\ 
	\hspace{50pt}  +\mu ^2\big( \ha \h ^2 \gi _0 T
     g_0 T + \ha \gi _0 T g_0 \h ^2 T-\h \gi _0 T g_0 \h T \big)    \bigg.
	\\ \hspace{50pt} + \ha \de \Psi_{_R} \commut{T}{\dm \de \Psi_{_R}+
	\commut{A_{0-}}{\de \Psi_{_R}}} 
	+\ha \de \Psi_{_L} \commut{T}{\dpl \de \Psi_{_L}+\commut{A_{0+}}{\de \Psi_{_L}}}
	 +\mu \gi_0 \de \Psi_{_L} g_0\de \Psi_{_R} \bigg] \,. 
	\ea
\end{equation}
This Lagrangian was derived in \cite{hit}.

In order to study fluctuations around a particular classical solution by using the reduced theory we need to find corresponding $g_0$, $A_{0\pm}$ by fixing the $G$ gauge and partially fixing the $H \times H$ gauge. 
Generally it is not easy to find a gauge such that $g_0$, $A_{0\pm}$ solve the gauge equations \eqref{redgaugef} as well as they take a convenient form for extracting physical part of the perturbation. 
Moreover it becomes much harder if the classical string solution is inhomogeneous, which is the case we will discuss in this paper. 
One can circumvent this gauge fixing task by using an embedding of the complex sinh-Gordon model into the deformed gWZW model.

Hence we shall next show how the complex sinh-Gordon model are realized in the context of Pohlmeyer reduction of the ${\rm AdS}_5 \times {\rm S}^5$ superstring theory.  
Because we are now interested in classical solutions whose ${\rm S}^5$ part is in ${\rm S}^1$ of the ${\rm S}^5$, 
the ${\rm S}^5$ parts of $g_0$ and $A_{0\pm}$ are the vacuum solution, and accordingly, 
the bosonic fluctuations in the ${\rm S}^5$ sector are massive fields with the masses $\pm\mu$ as shown in \cite{hit}.
Hence we will focus on the AdS sector, then $g_0$ and $A_{0\pm}$ below always mean matrices for the AdS sector.

A classical string solution in ${\rm AdS}_3 \times {\rm S}^1$ can be expressed as a classical solution $g_0$ which takes value on $SU(1,1)$ in the gWZW model. 
One natural parameterization of $su(1,1)$ as a subalgebra of $sp(2,2)$ is 
\be  \ba{c}
R_1=\left(
\begin{array}{cccc}
 0 & 0 & 0 & 1 \\
 0 & 0 & 1 & 0 \\
 0 & 1 & 0 & 0 \\
 1 & 0 & 0 & 0
\end{array}
\right) \,,~~~
R_2=\left(
\begin{array}{cccc}
 i & 0 & 0 & 0 \\
 0 & -i & 0 & 0 \\
 0 & 0 & i & 0 \\
 0 & 0 & 0 & -i
\end{array}
\right) \,,~~~
R_3=\left(
\begin{array}{cccc}
 0 & 0 & 0 & i \\
 0 & 0 & -i & 0 \\
 0 & i & 0 & 0 \\
 -i & 0 & 0 & 0
\end{array}
\right) \,.
\ea \ee
Then the classical solution is expressed in terms of the Euler angles, $\phi_{\!_A} $ and $\chi_{\!_A}$,
\begin{equation}
\bar{g}_0=g_2g_1g_2\,, ~~~g_1=\exp(\phi_{\!_A} R_1)\,,~~~g_2=\exp(\ha \chi_{\!_A} R_2)\,,
\end{equation}
and the matrix elements of $\bar{g}_0$ are written as, 
\begin{equation} \label{gcoth}
\bar{g}_0=\left(
\begin{array}{cccc}
 e^{i \chi_{\!_A}} \cosh\phi_{\!_A} & 0 & 0 & \sinh\phi_{\!_A} \\
 0 & e^{-i \chi_{\!_A}} \cosh\phi_{\!_A} & \sinh\phi_{\!_A} & 0 \\
 0 & \sinh\phi_{\!_A} & e^{i \chi_{\!_A}} \cosh\phi_{\!_A} & 0 \\
 \sinh\phi_{\!_A} & 0 & 0 & e^{-i \chi_{\!_A}} \cosh\phi_{\!_A}
\end{array}
\right)\,.
\end{equation}
The corresponding gauge fields are obtained by solving the gauge field equations \eqref{redgaugef},
\be
	\bar{A}_{0\pm }=\frac{i}{2} \, \bar{a}_\pm R_2 \,, 
\ee 
where
\be \label{gacot}\ba{c}
\bar{a}_+=- \text{coth}^2 \phi_{\!_A}  \partial _+ \chi_{\!_A}  \,, \\
  \bar{a}_-=  \text{coth}^2 \phi_{\!_A}  \partial _- \chi_{\!_A} \,.
\ea\ee
Plugging these into the gWZW Lagrangian in \cite{gt1}, one finds that the Lagrangian of the coth model, \eqref{comschc}, is recovered.

On the other hand the Lagrangian of the tanh model is obtained if we choose the following parameterization of $g_0$, 
 \be \label{g0ta} \ba{c}g_0 =
\left(
\begin{array}{cccc}
 0 & e^{i \theta_{\!_A}} \text{cosh}\phi_{\!_A} & -e^{i \theta_{\!_A}} \text{sinh}\phi_{\!_A} & 0 \\
 -e^{-i \theta_{\!_A}} \text{cosh}\phi_{\!_A} & 0 & 0 & e^{-i \theta_{\!_A}} \text{sinh}\phi_{\!_A} \\
 e^{i \theta_{\!_A}} \text{sinh}\phi_{\!_A} & 0 & 0 & -e^{i \theta_{\!_A}} \text{cosh}\phi_{\!_A} \\
 0 & -e^{-i \theta_{\!_A}} \text{sinh}\phi_{\!_A} & e^{-i \theta_{\!_A}} \text{cosh}\phi_{\!_A} & 0
\end{array}
\right)\,.
\ea \ee
The gauge equations \eqref{redgaugef} are solved by
\be
	A_{0\pm } =\frac{i}{2}\,a_\pm R_2 \,, 
\ee 
where 
\be \label{gatan}\ba{c}
a_+=-2\frac{ \text{cosh}2\phi_{\!_A} \, \partial_+\theta_{\!_A}}{1+\text{cosh}2\phi_{\!_A}} \,, \\
  a_-=- \text{sech}^2\phi_{\!_A} \, \partial_-\theta_{\!_A} \,.
\ea\ee
Note that $\bar{g}_0$ and $g_0$ are related by an $H \x H$ gauge transformation rather than an $H$ gauge transformation. One example of the $H$ gauge transformation is 
\be
 \bar{g}_0 \to g_0= h_L^{-1}\bar{g}_0 h_R\,,
\ee
where 
\be
h_L=
\left(
\begin{array}{cccc}
 0 & -e^{\frac{3}{2} i \theta_{\!_A}} & 0 & 0 \\
 e^{-\frac{3}{2} i\text{  }\theta_{\!_A}} & 0 & 0 & 0 \\
 0 & 0 & 0 & -e^{\frac{3}{2} i \theta_{\!_A}} \\
 0 & 0 & e^{-\frac{3}{2} i \theta_{\!_A}} & 0
\end{array}
\right)\,, ~~~~~
 h_R=\left(
\begin{array}{cccc}
 e^{-\frac{1}{2} i \theta_{\!_A}} & 0 & 0 & 0 \\
 0 & e^{\frac{1}{2} i \theta_{\!_A}} & 0 & 0 \\
 0 & 0 & -e^{-\frac{1}{2} i \theta_{\!_A}} & 0 \\
 0 & 0 & 0 & -e^{\frac{1}{2} i \theta_{\!_A}}
\end{array}
\right) \,.
\ee
Because the corresponding gauge fields \eqref{gacot} and \eqref{gatan} are not necessarily connected by the $H \x H$ gauge transformation, 
we should resolve the gauge equations \eqref{redgaugef} once we derive the new $g_0$.

Below we will evaluate the fluctuations in the whole  ${\rm AdS}_5 \times {\rm S}^5$ around classical strings stretching in the  ${\rm AdS}_3 \times {\rm S}^1$ subspace by 
using the embedding of the complex sinh-Gordon model into the deformed gWZW model, i.e., \eqref{g0ta} and \eqref{gcoth}. 
Although $g_0$ for the coth model is constructed by a standard gauging,  the coth model has several issues when we calculate quadratic fluctuations, and we 
will mainly use the embedding of the tanh model.   
We will come back to this point in the case of the $(S,J)$ folded string in the next section.

\renewcommand{\theequation}{3.\arabic{equation}}
 \setcounter{equation}{0}

\section{Folded string\label{secfold}}

In this section we shall evaluate the Lagrangian for quadratic fluctuations around the $(S,J)$ folded string. 
In the original string theory the semiclassical expansions are carried out in the Nambu action in the static gauge and in the Polyakov action in the conformal gauge in \cite{ft1}, and the equivalence of these two approaches is shown in \cite{bdfpt}. 
Although the Pohlmeyer reduced form of the ${\rm AdS}_5 \times {\rm S}^5$ is constructed from the conformal gauge string theory, it is expected that the fluctuation Lagrangian of the reduced theory takes a similar form to the effective Nambu action as both the reduced model and  the Nambu action involve only the physical degrees of freedom after choosing a gauge. 

We shall derive the Lagrangian for bosonic fluctuations in \ref{redfold} and the Lagrangian for fermionic fluctuations in \ref{foldfer}. 
To compare our result with the original string theory we shall carry out the perturbation in the Nambu action in the static gauge, and show how this approach is related to the coth model and tanh model in the reduced theory in \ref{foldsta}. 

Let us first review the $(S,J)$ folded string in ${\rm AdS}_3 \times {\rm S}^1$ which is expressed in terms of the embedding coordinates, 
\begin{equation}\label{gkpt} \ba{c}
Y_0+iY_{3}=\cosh \! \rho  \,e^{i\kappa \tau}\,, 
~~~ Y_1+iY_2= \sinh \! \rho \,e^{i w \tau}\,,  ~~~ 
X_1+iX_2=e^{i\nu \tau}\,,
\ea
\end{equation}
where $\kappa$, $w$ and $\nu$ are constants, and $\rho$ is a function of $\sigma$, $\rho=\rho(\sigma)$. 
The equation of motion and the conformal gauge constrains read 
\begin{equation}\label{gkptr} \ba{c}
\rho^{\prime \prime}=(\kappa^2-w^2)\sinh \! \rho \, \cosh \! \rho \,, \\
\rho^{\prime 2 }=\kappa^2 \cosh^2 \! \rho -w^2 \sinh^2 \! \rho -\nu^2 \, .
\ea
\end{equation}
 
Next we shall derive the corresponding classical solution in the reduced theory. 
The mass scale of the reduced theory $\mu$ can be found by observing the AdS part of the stress tensor. In the present case we have $T_{\pm \pm}^{\rm AdS}=-\nu^2$ meaning that this is the case of $\mu_+=\mu_-$. Then we set 
\be
\mu=\nu \,.
\ee
We have two ways of the reduction, the embedding of the coth model or the embedding of the tanh model. 
As mentioned below it is more convenient to employ the tanh approach. 
The classical solution, $\phi_{\!_A}$ and $\theta_{\!_A}$, in the tanh model is given by the relations in \eqref{red}, 
\be \label{soltanh}\ba{c}
\phi_{\!_A}=\text{log}\left( \frac{\rho^{\prime}+\sqrt{\nu ^2+\rho^{\prime 2}}}{\nu }\right) \,, \\
\theta_{\!_A} =\frac{ w \kappa }{\nu }\tau  \,.
\ea\ee
Substituting these into the formula of $g_0$ and $A_{\pm}$, we have 
\be \label{gfold} g_0=
\left(
\begin{array}{cccc}
 0 & v\frac{ \sqrt{\nu ^2+\rho^{\prime 2}}}{\nu } & -v\frac{ \rho^{\prime}}{\nu } & 0 \\
 -v^*\frac{ \sqrt{\nu ^2+\rho^{\prime 2}}}{\nu } & 0 & 0 & v^*\frac{ \rho^{\prime}}{\nu } \\
 v\frac{ \rho^{\prime}}{\nu } & 0 & 0 & -v\frac{ \sqrt{\nu ^2+\rho^{\prime 2}}}{\nu } \\
 0 & -v^*\frac{ \rho^{\prime}}{\nu } & v^*\frac{ \sqrt{\nu ^2+\rho^{\prime 2}}}{\nu } & 0
\end{array}
\right) \,, ~~~v=e^{\frac{i  w \kappa \tau }{\nu }} \,,
\ee 
and the gauge field equations \eqref{redgaugef} are solved by the following $A_{\pm 0}$,
\be
	A_{\pm 0}=\frac{i}{2}\,a_{\pm 0} R_2 \,, 
\ee 
where $a_{\pm 0}$ are given by 
\be\ba{c}
a_{+0}=2 w \kappa  \left(-\frac{1}{\nu }+\frac{\nu }{2 \left(\nu ^2+\rho^{\prime 2}\right)}\right) \,, \\
  a_{-0}=-\frac{ w \kappa  \nu }{ \left(\nu ^2+\rho^{\prime 2}\right)} \,.
\ea\ee

The reason we use the tanh model here is as follows. 
If we plug the folded string solution \eqref{gkpt} into the reduction equation \eqref{redcoth} with $\mu =\nu $, we obtain the classical solution of the coth model 
\be \label{solcoth} \ba{c}
\phi_{\!_A}=\text{log}\left( \frac{\rho^{\prime}+\sqrt{\nu ^2+\rho^{\prime 2}}}{\nu }\right) \,, \\
  \partial_\pm \chi_{\!_A} =\mp \left( \frac{w \kappa}{\nu }-\frac{w \kappa \nu }{\nu ^2+\rho '^2} \right)\,.
\ea \ee
Here $\chi_{\!_A}$ can be expressed in an integral form. This can be also understood by substituting $\phi_{\!_A}$ into the relation of $\chi_{\!_A}$ and $\theta_{\!_A}$ in \eqref{tdual}. 
On the right hand side, $\partial_\pm \theta_{\!_A}$ is constant, but $\tanh \phi_{\!_A}$ has the $\rho(\sigma)$ dependence, and then, $\chi_{\!_A}$ ends up with the complicated form.

As far as the calculation of the quadratic fluctuations is concerned, 
the expression of $\chi_{\!_A}$ in \eqref{solcoth} does not seem to cause a serious problem 
because only its derivative, $\partial_+\chi_{\!_A}$, appears in to the Lagrangian for the quadratic fluctuations if the $H\x H$ gauge is properly chosen. 
However, $\bar{g}_0$ for the coth model in \eqref{gcoth} is not this case; due to the peculiar form of $\bar{g}_0$ 
some components of $\bar{g}_0^{-1}\partial_+\bar{g}_0$ and $\bar{g}_0^{-1}\bar{A}_{0+}\bar{g}_0$ contain the $e^{i\chi_{\!_A}}$ factor, and consequently, 
the perturbed Lagrangian has the $\chi_{\!_A}$ dependence. 
Of course this $e^{i\chi_{\!_A}}$ factor can be removed from the perturbed Lagrangian by redefinition of fluctuation fields, but such redefinition is highly nontrivial. 
Also, the form of $g_0$ for the tan model in \eqref{g0ta} is very convenient to integrate out the gauge fields and decouple the physical fields from the unphysical fields. 
Therefore we will basically use the tanh model in the following sections. 

Before moving on to the semiclassical computation, let us add some remarks on the open string counterpart of the folded string solution in the scaling limit where the $(S,J)$ folded string solution becomes homogeneous. 
The $(S,J)$ folded string in this limit was studied in the reduced theory \cite{hit}. 
In \cite{krtt} it was shown that the two classical string solutions are connected by the $SO(2,4)$ rotation and analytic
continuation on the worldsheet $\tau \to -i\tau$, and then, quantum corrections to the scaling function calculated in the open string picture are 
the same as those in the folded string picture.\footnote{
In \cite{krtt} the equivalence of the closed string and the open string was shown for $\nu=0$ classically and semiclassically. 
A $\nu \ne 0$ open string solution can be easily constructed by adding the BMN vacuum in the ${\rm S}^1$ sector \cite{rotecusp}, which is the counterpart of the scaling limit of the $(S,J)$ folded string (see also \cite{grrtv} for a folded string and its corresponding open string surface with both the orbital momentum and the winding in ${\rm S}^1$ of ${\rm S}^5$).  
}
Because the isometry of ${\rm AdS}_5$, $SO(2,4)$, becomes obscure by the Pohlmeyer reduction, 
any two solutions related by an $SO(2,4)$ transformation are encoded into a single solution in the reduced theory.
Hence we obtain a reduced theory solution corresponding to the null cusp solution in the original theory by the analytic
continuation $\tau \to -i\tau$ in $g_0$ \eqref{gfold} (after taking the scaling limit, $w \to \kappa$ and $\rho \to \ell \sigma$ with $\kappa^2 - \nu^2= \ell^2$). 
In general the equivalence of characteristic frequencies of the quadratic fluctuations in the reduced theory comes from that of the classical solutions $g_0$. 
Therefore, in the reduced theory, one can trivially check frequencies for the two classical solutions match.

\subsection{Bosonic fluctuations in reduced theory\label{redfold}}

The argument for the ${\rm S}^5$ sector is the same as the case of the scaling limit studied in \cite{hit}. 
Four bosonic fluctuations in the ${\rm S}^5$ sector are massive fields with $m_B^2=\nu^2$. 

In the ${\rm AdS}_5$ sector we shall discuss the bosonic fluctuations by using the tanh model for the above reason. 
To express the quadratic fluctuations in terms of components fields, let us introduce bosonic fields by
\be\label{etap}
	\h^\parallel =\left(
	\begin{array}{cccc}
		0 & 0 & a_{1}+ia_{2}& a_{3}+ia_{4}\\
		0 & 0 & a_{3}-ia_{4}& -a_{1}+ia_{2}\\
		a_{1}-ia_{2}& a_{3}+ia_{4}& 0 & 0 \\
		a_{3}-ia_{4}& -a_{1}-ia_{2}& 0 & 0  
	\end{array}
	\right)\,,
\ee
which correspond to physical fields in the reduced theory. 
\be\label{etao} 
\h^\perp=\left(\ba{cccc}
i h_1 & h_2+i h_3&0&0
\\-h_2+i h_3&-i h_1&0&0
\\0&0&i h_4 & h_5+i h_6
\\0&0&-h_5+i h_6&-i h_4\ea\right)\, ,\ee
\be \label{dgau} \ba{c}
\delta A_+=\left(\ba{cccc}
i a_{+1} & \left(a_{+2}+i a_{+3}\right)v^2&0&0
\\-\left(a_{+2}-i a_{+3}\right)v^{*}{}^2&-i a_{+1}&0&0
\\0&0&i a_{+4} & \left(a_{+5}+i a_{+6}\right)v^2
\\0&0&-\left(a_{+5}-i a_{+6}\right) v^{*}{}^2&-i a_{+4}\ea\right)\, ,
\\
\delta A_-=\left(\ba{cccc}
i a_{-1} & a_{-2}+i a_{-3}&0&0
\\-a_{-2}+i a_{-3}&-i a_{-1}&0&0
\\0&0&i a_{-4} & a_{-5}+i a_{-6}
\\0&0&-a_{-5}+i a_{-6}&-i a_{-4}\ea\right)\, .\ea\ee
These are unphysical fields in the reduced theory and to be gauged away or integrated out from the fluctuation Lagrangian.

One advantage of using $g_0$ \eqref{g0ta} is that the system decouples into two subsectors: 
One contains $a_1$ and $a_2$ coupling to the diagonal parts of $\h^\perp$ and $\delta A_\pm$, 
while the other contains $a_3$ and $a_4$ coupling to the off-diagonal components of $\h^\perp$ and $\delta A_\pm$. 
Hence, as done in the long string limit case in \cite{hit}, we can fix the $H$ gauge such that the physical fields $a_i$ decouple from the unphysical fields, 
\be
h_1+h_4=\trm{const}\,, 
\ee 
which is the same as one of the three gauge conditions in the long string limit case in \cite{hit}, and we should impose two more conditions, 
\be
\ba{c}
\sqrt{\left(-w^2+\nu ^2+\rho'^2\right) \left(-\kappa ^2+\nu ^2+\rho'^2\right)} \hspace{270pt} \\ 
\times \bigg[ w \kappa  \nu ^3 (h_3+h_6)  +\left(\nu ^2+\rho'^2\right) \Big( \left(\nu ^2+2 \rho'^2\right) (a_{+2}+a_{+5})  +\nu ^2 \left(\partial_-h_2+\partial_-h_5 -a_{-2}-a_{-5} \right)\Big) \bigg]  \\
+\rho' \bigg[ -w^2 \kappa ^2 \nu ^2 (h_2+h_5)+\left(\nu ^2+\rho'^2\right) \Big( w \kappa  \nu  \left(\partial_-h_3+\partial_-h_6-a_{-3}-a_{-6}\right) \hspace{100pt} \\
\hspace{210pt} +\left(\nu ^2+\rho'^2\right) \left(\partial_-a_{+2}+\partial_-a_{+5}\right)\Big) \bigg]=0\,, 
\ea\ee
and
\be\ba{c}
\sqrt{\left(-w^2+\nu ^2+\rho'^2\right) \left(-\kappa ^2+\nu ^2+\rho'^2\right)} \hspace{270pt} \\ 
\times \bigg[ w \kappa  \nu ^3 (-h_2+h_5)  -\left(\nu ^2+\rho'^2\right) \Big( (a_{+3}-a_{+6}) \left(\nu ^2+2 \rho'^2\right) +\nu ^2 \left(-\partial_-h_3+\partial_-h_6+a_{-3}-a_{-6}\right)\Big) \bigg] \\ 
+\rho' \bigg[ w^2 \kappa ^2 \nu ^2 (-h_3+h_6)+\left(\nu ^2+\rho'^2\right) \Big( w \kappa  \nu  \left(-\partial_-h_2+\partial_-h_5+a_{-2}-a_{-5}\right) \hspace{100pt} \\
\hspace{210pt} -\left(\nu ^2+\rho'^2\right) \left(\partial_-a_{+3}-\partial_-a_{+6}\right)\Big) \bigg] =0\,.
\ea\ee 
Under this gauge choice the Lagrangian for a sector with $a_1$ and $a_2$ is 
\be \label{ell1}
\mathcal{L}_1=2\sum _{i=1,2}\left( \partial_+a_i \partial_- a_i -\left(\nu ^2+2 \rho ^{\prime 2} \right)a_i^2\right)\,,
\ee
and the sector with $a_3$ and $a_4$ has the Lagrangian,
\be \label{ell2} \ba{c}
\mathcal{L}_2= 2\bigg[ \partial _- a_3 \partial _+ a_3
-\left( \nu ^2+2 \rho '^2+\frac{2 w^2 \kappa ^2}{\nu ^2+\rho '^2}-\frac{3 w^2 \kappa ^2 \nu ^2}{\left(\nu ^2+\rho '^2\right)^2}\right) a_3 ^2 +\partial _- a_4 \partial _+ a_4 
\\
~~~~~~~~~~~~+\left(\frac{2 w \kappa  \nu  \partial _- a_3}{\nu ^2+\rho ^{\prime 2}}+\frac{2 w \kappa  \nu  \partial _+ a_3}{\nu ^2+\rho ^{\prime 2}}\right) a_4
- \nu ^2 \left(-1+\frac{2 \left(w^2+\kappa ^2\right) }{\left(\nu ^2+\rho '^2\right)}-\frac{3 w^2 \kappa ^2}{\left(\nu ^2+\rho '^2\right)^2}\right)  a_4 ^2 
\bigg]
 \,.
\ea\ee
This Lagrangian does not diverge at turning points of the folded string, i.e., at $\rho'=0$. This observation is different from the case of $\nu = 0$ in \cite{ft1}.
The long string limit case is recovered where we take $w \to \kappa$ and $\rho \to \ell \sigma$, and replace the conformal gauge 
constraint by $\kappa^2 - \nu^2= \ell^2$. In this limit the resulting Lagrangian yields the correct frequencies \cite{ftt}.

If we take $\nu \to 0$ limit in \eqref{ell1} and \eqref{ell2}, the Lagrangians become
\be \label{a1a2nu0}
\mathcal{L}_1=2\sum _{i=1,2}\left( \partial_+a_i \partial_- a_i -2 \rho ^{\prime 2} a_i^2\right)\,,
\ee
and
\be \label{a3a4}\ba{c}
\mathcal{L}_2=  2\left[ \partial _- a_3 \partial _+ a_3-\frac{2 \left(w^2 \kappa ^2+\rho ^{\prime 4}\right)}{ \rho ^{\prime 2}} a_3 ^2+\partial _- a_4 \partial _+ a_4 \right]
\,.
\ea\ee
The sum of these two Lagrangians are the exactly the same as the Lagrangian (5.6) in \cite{ft1}, which is found by perturbing the Nambu action in the static gauge. 
We find $a_1$ and $a_2$ correspond to $\beta_i$ while $a_3$ corresponds to $\phi_{\!_A}$ and 
$a_4$ is interpreted as a massless fluctuation denoted as $\tilde{\varphi}$ in \cite{ft1}.

So far we have carried out the perturbation in the full reduced theory by embedding the tanh model into the gWZW model and shown that the perturbed Lagrangian takes 
the Nambu-Goto type in the original string theory. 
One may expect two of the fluctuations are captured by perturbing the tanh model directly. 
Plugging the folded string solution \eqref{soltanh} into the fluctuation Lagrangian \eqref{quadtanh} and rescaling $\delta\theta_{\!_A}$ by 
\be
 \delta \theta_{\!_A} \to \frac{ \delta \theta_{\!_A}}{\sqrt{1-\frac{\nu ^2}{\nu ^2+\rho'^2}}},
\ee
then we have the following Lagrangian, 
\be \label{tanfol}\ba{c}
\mathcal{L}_{\rm tanh}= 
\partial _- \delta \phi_{\!_A} \partial _+ \delta \phi_{\!_A}
-\left( \nu ^2+2 \rho '^2+\frac{2 w^2 \kappa ^2}{\nu ^2+\rho '^2}-\frac{3 w^2 \kappa ^2 \nu ^2}{\left(\nu ^2+\rho '^2\right)^2}\right) \delta \phi_{\!_A} ^2 +\partial _- \delta \theta_{\!_A} \partial _+ \delta \theta_{\!_A} 
\\
~~~~~~~~~~~~+\left(\frac{2 w \kappa  \nu  \partial _-  \delta \phi_{\!_A} }{\nu ^2+\rho ^{\prime 2}}+\frac{2 w \kappa  \nu  \partial _+  \delta \phi_{\!_A} }{\nu ^2+\rho ^{\prime 2}}\right) \delta \theta_{\!_A}
- \nu ^2 \left(-1+\frac{2 \left(w^2+\kappa ^2\right) }{\left(\nu ^2+\rho '^2\right)}-\frac{3 w^2 \kappa ^2}{\left(\nu ^2+\rho '^2\right)^2}\right)  \delta \theta_{\!_A} ^2  \,.
\ea\ee
Noticing that this Lagrangian takes the same form as \eqref{ell2}, we find that $\delta \phi_{\!_A}$ and $\delta \theta_{\!_A}$ correspond to $a_3$ and $a_4$, respectively.
Hence it turns out that the perturbation in the tanh model describes the most complicated part of the bosonic fluctuation Lagrangian. 
This fact is useful when we consider more complicated classical solutions (see sections \ref{sjcs}, \ref{sjspik} and \ref{mwfold}).

\subsection{Fermionic fluctuations\label{foldfer}}
For a consistency with the original string theory, the masses of the fermionic fluctuations in the reduced theory should match those in the original theory. 
We define component fields of the fermionic fluctuations in the following way, 
\begin{gather}\label{ferRL1}
	\de \Psi_{_R}=\left(
	\begin{array}{cc}
		0& \mf{X}_R \\
		\mf{Y}_R& 0 
	\end{array}
	\right) \,, ~~~
	\de \Psi_{_L}=\left(
	\begin{array}{cc}
		0& \mf{X}_L  \\
		\mf{Y}_L& 0 
	\end{array}
	\right) \,,
\end{gather}
where
\begin{gather}
	\mf{X}_R=\left(
	\begin{array}{cccc}
		 0 & 0 &\alpha_1+i\alpha_2 &\alpha_3+i\alpha_4 \\
		 0 & 0 &-\alpha_3+i\alpha_4 &\alpha_1-i\alpha_2 \\
		\alpha_5+i\alpha_6 &\alpha_7-i\alpha_8 &0&0\\
		\alpha_7+i\alpha_8 &-\alpha_5+i\alpha_6 &0 &0 \\
	\end{array}
	\right)\,, \\
	\mf{Y}_R=\left(
	\begin{array}{cccc}
		0 & 0 &-\alpha_6 -i\alpha_5 & -\alpha_8 -i\alpha_7 \\
		0 & 0 &\alpha_8 -i\alpha_7 & -\alpha_6 +i\alpha_5 \\
		\alpha_2 +i\alpha_1 & \alpha_4 -i\alpha_3 &0 &0  \\
		\alpha_4 +i\alpha_3 & -\alpha_2 +i\alpha_1&0 &0  
	\end{array}
	\right)\,,
\end{gather} 
and
\small
\begin{gather}
	\mf{X}_L=\left(
	\begin{array}{cccc}
		0 & 0 &(\beta_1+i\beta_2)v  &(\beta_3+i\beta_4)v  \\
		0 & 0 &(\beta_3-i\beta_4 )v^*  &(-\beta_1+i\beta_2)v^* \\
		(\beta_5+i\beta_6 )v  &(-\beta_7+i\beta_8)v &0&0\\
		(\beta_7+i\beta_8 )  v^*  &(\beta_5-i\beta_6)v ^*  &0 &0 
	\end{array}
	\right)\,,   \\
	\mf{Y}_L=\left(
	\begin{array}{cccc}
		0 & 0 &(-\beta_6 -i\beta_5)v^* &( -\beta_8 -i\beta_7)v \\
		0 & 0 &(-\beta_8 +i\beta_7 )v^*  &( \beta_6 -i\beta_5)v  \\
		(\beta_2 +i\beta_1)v^* & (-\beta_4 +i\beta_3)v &0 &0  \\
		(\beta_4 +i\beta_3)v^* & (\beta_2 -i\beta_1 )v &0  &0  
	\end{array}
	\right)\,, \label{ferRL5}
\end{gather}
where all component fields are real Grassmann. 
The extra factor $v=\exp(iw \kappa \ta/\nu)$ is introduced in $\mf{X}_L$ and $\mf{Y}_L$ such that the exponential factor does not appear in the Lagrangian. 
The resulting Lagrangian is 
\be\ba{c}
\mathcal{L}_{\rm F}=2\bigg[\sum\limits_{i=1}^8 \left( \alpha_i \partial_-\alpha_i +\beta_i \partial_+\beta_i\right) \hspace{160pt} \\ \hspace{23pt}
+\frac{w \kappa  \nu }{\nu ^2+\rho '^2}\left( \alpha_1\alpha_2+\alpha_3\alpha_4+\alpha_5\alpha_6-\alpha_7\alpha_8+\beta_1\beta_2+\beta_3\beta_4+\beta_5\beta_6-\beta_7\beta_8 \right) \\ \hspace{56pt}
+2 \sqrt{\nu ^2+\rho '^2} \left( -\alpha_3\beta_1  -\alpha_1\beta_3  +\alpha_7\beta_5  -\alpha_5\beta_7  +\alpha_4\beta_2  +\alpha_2\beta_4 + \alpha_8\beta_6 - \alpha_6\beta_8   \right)
\bigg] \,,
\ea\ee
After a careful observation we find this system decouples into four subsectors, which show an identical structure containing two of $\alpha_i$ fields and two of $\beta_i$ fields. 
Let us focus on the subsector containing $\alpha_1$, $\alpha_2$, $\beta_3$ and $\beta_4$, described by a part of the Lagrangian, 
\be\ba{c}
\mathcal{L}= \alpha_1 \partial_-\alpha_1 +\alpha_2 \partial_-\alpha_2 +\beta_3 \partial_+\beta_3+\beta_4 \partial_+\beta_4
+\frac{w \kappa  \nu }{\nu ^2+\rho '^2}\left( \alpha_1\alpha_2+\beta_3\beta_4 \right)
+2 \sqrt{\nu ^2+\rho '^2} \left(  -\alpha_1\beta_3   +\alpha_2\beta_4   \right)
 \,,
\ea\ee
which is simplified as 
\be \label{fersub}
\mathcal{L}= \bar\psi \gamma^a \partial_a\psi + \ha \frac{w \kappa  \nu }{\nu ^2+\rho '^2} \bar\psi\Gamma_1\psi - \sqrt{\nu ^2+\rho '^2}\bar\psi\Gamma_2\psi \,,
\ee
where 
\be
\psi =\left( 
\ba{c}
\beta_3 \\ \beta_4 \\
\alpha_1 \\ \alpha_2
\ea \right) \,,~~~
\gamma^\tau =\left(
\begin{array}{cccc}
 0 & 0 & 0 & -i \\
 0 & 0 & -i & 0 \\
 0 & i & 0 & 0 \\
 i & 0 & 0 & 0
\end{array}
\right) \,,~~~
\gamma^\sigma =\left(
\begin{array}{cccc}
 0 & 0 & 0 & i \\
 0 & 0 & i & 0 \\
 0 & i & 0 & 0 \\
 i & 0 & 0 & 0
\end{array}
\right) \,,~~~
\bar\psi= \psi^{\dagger}\gamma^{\tau} \,,
\ee
and
\be
\Gamma_1 =\left(
\begin{array}{cccc}
 0 & 0 & i & 0 \\
 0 & 0 & 0 & -i \\
 -i & 0 & 0 & 0 \\
 0 & i & 0 & 0
\end{array}
\right) \,,~~~
\Gamma_2 =\left(
\begin{array}{cccc}
 0 & i & 0 & 0 \\
 -i & 0 & 0 & 0 \\
 0 & 0 & 0 & i \\
 0 & 0 & -i & 0
\end{array}
\right) \,,
\ee
The first two terms in \eqref{fersub} can be summarized by introducing $\Psi\equiv \exp \left(  a \Gamma_3 \right)\psi$ with 
\be
\frac{da}{d\sigma}=\ha \frac{w \kappa  \nu }{\nu ^2+\rho '^2} \,, ~~\Gamma_3  =
\left(
\begin{array}{cccc}
 0 & 1 & 0 & 0 \\
 -1 & 0 & 0 & 0 \\
 0 & 0 & 0 & -1 \\
 0 & 0 & 1 & 0
\end{array}
\right)\,.
\ee
Using that $\Gamma_3^{\dagger}\gamma^\tau\gamma^\sigma \Gamma_3=\gamma^\tau\gamma^\sigma$ and $\Gamma_3^{\dagger}\gamma^\tau\Gamma_2 \Gamma_3=\gamma^\tau\Gamma_2$
then we find the Lagrangian \eqref{fersub} becomes 
\be \label{ferpfld}
\mathcal{L}_{\rm F}= \bar\Psi \gamma^a \partial_a\Psi  - \sqrt{\nu ^2+\rho '^2}\bar\Psi\Gamma_2\Psi \,.
\ee
The same prescription applies for the other subsectors, and consequently, 
it turns out that the fermionic fluctuations have the mass term with a $\sigma$-dependent coefficient $\sqrt{\nu ^2+\rho '^2}$, which agrees with \cite{ft1}.

\subsection{Bosonic fluctuations from Nambu action of original string theory\label{foldsta}}

In \cite{ft1} the authors showed the bosonic fluctuations around the folded string without the $S^1$ part ($\nu=0$ in \eqref{gkpt})
and they found that the mass of one bosonic fluctuation found from the Nambu action contains one $1/\rho'^2 $ term. 
However, our calculations in the reduced theory in \ref{redfold} shows that the Lagrangian has no $1/\rho'^2$ term 
if the string solution has the ${\rm S}^1$ sector. 

It might be considered that this difference appears because the fluctuations in the ${\rm AdS}_5$ part nontrivially couple to a fluctuation in ${\rm S}^5$ part in the $\nu \ne 0$ case.
So we shall first carry out the perturbation in the Nambu action with nonvanishing $\nu$. 
However we will find the nontrivial coupling is not all of the reason for the difference; 
the fluctuations in the tanh model and in the Nambu action are related by the T-duality transformation rather than 
rescaling or rotation of fluctuation fields. 
Because the partition functions of any two theories connected by the T-duality transformation are the same, 
this is a nontrivial support for our conjecture on the quantum equivalence \eqref{eqpa}. 

In the reduced theory the T-dual of the tanh model is the coth model. So we shall also show that the perturbation in the coth model recovers the quadratic fluctuations found by perturbing the Nambu action in another gauge. In section \ref{tancotwzw} we showed that the tanh model and coth model are realized as the different ways of $H \x H$ gauge fixing. 
Hence the result supports that the partition function is $H \x H$ gauge independent.

The Nambu action for the bosonic string in ${\rm AdS}_5 \times {\rm S}^5$ is given by 
\be \label{nago}
	S_{\rm N} =- \int d\tau d\sigma \sqrt{-{\rm det} h_{ab}} \,,
\ee
where the induced metric on the worldsheet $h_{ab}$ is 
\be
	h_{ab}=g_{\mu \nu}(x)\partial_a x^\mu \partial_b x^{\nu} \,.
\ee
In the present case the classical forms of $g_{\mu \nu}(x)$ and $x^\mu$ are respectively
\be\ba{c}
	g_{\mu \nu}(x)dx^\mu dx^\nu =-\cosh ^2 \! \rho \, dt^2+ d\rho^2+ \sinh ^2\! \rho \left( d\beta^2_i + d\phi^2  \right) + d\psi _s^2+ d\varphi^2 ~~~~
(i=1,2,~s=1,2,3,4)\,,\\
	t=\kappa \tau \,,~~\rho=\rho(\sigma)\,,~~\beta_i=0\,,~~\phi=w\tau\,,~~\psi_s=0\,,~~\varphi=\nu \tau\,,
\ea\ee
which are related by the Virasoro constraints and equation of motion \eqref{gkptr}. 
Imposing the static gauge where the fluctuations of $t$ and $\rho$ are set to zero we have the following perturbation, 
\be\ba{c}
	t=\kappa \tau\,,~~\rho=\rho(\sigma)\,,~~ \beta_i=\frac{1}{\lambda^{1/4}}\tilde \beta_i \,,~~~ \phi = w \tau + \frac{1}{\lambda^{1/4}}\tilde{\phi}\,,~~
	\psi_s=\frac{1}{\lambda^{1/4}}\tilde\psi_s\,,~~\varphi=\nu \tau + \frac{1}{\lambda^{1/4}}\tilde\varphi \,,
\ea
\ee
Expanding the Lagrangian \eqref{nago} gives the following action for the quadratic fluctuations, 
\be\label{ngper} \ba{rl}
& S_{{\rm N}} = \ha \int d\tau d\sigma \bigg[
\sinh ^2\! \rho \partial_+\tilde\beta _i \partial_- \tilde\beta_i- w^2 \sinh ^2\! \rho \tilde\beta_i^2 + \partial_+\tilde\psi_s \partial_- \tilde\psi_s- \nu ^2 \tilde\psi_s^2 \nonumber \\ & \hspace{45pt}
+ \sinh ^2\! \rho \left(1+\frac{w^2 \sinh ^2\! \rho}{\rho'^2}\right)  \partial_+\tilde\phi  \partial_-\tilde\phi +\left(1+\frac{\nu ^2}{ \rho'^2}\right)  \partial_+\tilde\varphi  \partial_-\tilde\varphi+\frac{w \nu  \sinh ^2\! \rho \, \left( \partial_+\tilde\varphi  \partial_-\tilde\phi+ \partial_+\tilde\phi  \partial_-\tilde\varphi\right)}{ \rho'^2} \bigg]\,. 
\ea
\ee
This shows the fluctuations $\tilde\psi_s$ have $m^2_{\tilde\psi_s}=\nu^2$. They describe the four fluctuations in the ${\rm S}^5$ sector in the reduced theory. 
By rescaling $\tilde\beta_i$ by $\tilde\beta_i\to \sinh ^{-1}\! \rho \,\tilde\beta_i$, we find $\tilde\beta_i$ have $m^2_{\tilde\beta_i}=\nu^2+2 \rho'^2$. 
Hence the fluctuations $\beta_i$ correspond to $a_1$ and $a_2$ in \eqref{ell1}. 

Let us focus on the other two fields, $\tilde\phi$ and $\tilde\varphi$, which should be compared with $a_3$ and $a_4$ in \eqref{ell2}. 
The corresponding part in \eqref{ngper} is
\begin{equation}\label{ng34}
 \mathcal{L}_{\rm N}= F_1  \partial_+\tilde\phi  \partial_-\tilde\phi +F_2 \partial_+\tilde\varphi  \partial_-\tilde\varphi+ F_3\left( \partial_+\tilde\varphi  \partial_-\tilde\phi+ \partial_+\tilde\phi  \partial_-\tilde\varphi\right)\,,
\end{equation}
where 
\begin{equation}
F_1= \sinh ^2\! \rho \left(1+\frac{w^2 \sinh ^2\! \rho}{\rho'^2}\right) \,,~~ F_2=1+\frac{\nu ^2}{ \rho'^2} \,,~~F_3=\frac{w \nu  \sinh ^2\! \rho \, }{ \rho'^2}\,,
\end{equation}
Even if we rescale $\tilde\phi$ and $\tilde\varphi$ such that the coefficients of their kinetic terms are one, 
the resulting Lagrangian does not match the Lagrangian in \eqref{ell2}.
In fact \eqref{ng34} and \eqref{ell2} are related by the T-duality transformation. 
Because the Lagrangian \eqref{ng34} only depends on the derivatives of the fields $\phi$ and $\varphi$,  
it is allowed to carry out the T-duality transformation at the level of the Lagrangian in the present case. 
In order to derive the T-dualized Lagrangian we first denote $\partial_\pm\tilde\phi$ by $\mathcal{A}_\pm$ and introduce a Lagrange multiplier $\tilde{x}$. 
Then the Lagrangian \eqref{ng34} is becomes
\begin{equation}
 \mathcal{L}_{\rm N} = F_1  \mathcal{A}_+ \mathcal{A}_- +F_2 \partial_+\tilde\varphi  \partial_-\tilde\varphi+ F_3\left( \partial_+\tilde\varphi  \mathcal{A}_-+ \mathcal{A}_+  \partial_-\tilde\varphi\right) + \tilde{x}(\partial_+\mathcal{A}_--\partial_-\cal{A}_+)\,.
\end{equation}
The Lagrange multiplier $\tilde{x}$ will become a new physical field in the T-dual picture. 
Integrating out $\mathcal{A}_+$ yields 
\begin{equation}
 \mathcal{L}_{\rm TN}=\frac{1}{ F_1}  \partial_+\tilde{x} \partial_-\tilde{x} +\left(F_2-\frac{F_3^2}{F_1}\right) \partial_+\tilde\varphi  \partial_-\tilde\varphi+ \frac{F_3}{F_1}\left( \partial_+\tilde{x}  \partial_-\tilde\varphi- \partial_+\tilde\varphi  \partial_-\tilde{x} \right)\,.
\end{equation}
Finally, rescaling $\tilde{x} \to F_1^{1/2}\tilde{x}$ and $\tilde\varphi \to \left(F_2-\frac{F_3^2}{F_1}\right)^{-1/2} \tilde\varphi$, 
we have the following Lagrangian, 
\be \ba{c}
\mathcal{L}_{\rm TN}= 
\partial _- \tilde\varphi \partial _+ \tilde\varphi
-\left( \nu ^2+2 \rho '^2+\frac{2 w^2 \kappa ^2}{\nu ^2+\rho '^2}-\frac{3 w^2 \kappa ^2 \nu ^2}{\left(\nu ^2+\rho '^2\right)^2}\right) \tilde\varphi ^2 +\partial _-  \tilde{x} \partial _+  \tilde{x}
\\
~~~~~~~~~~~~+\left(\frac{2 w \kappa  \nu  \partial _-  \delta \phi }{\nu ^2+\rho ^{\prime 2}}+\frac{2 w \kappa  \nu  \partial _+  \delta \phi }{\nu ^2+\rho ^{\prime 2}}\right)  \tilde{x}
- \nu ^2 \left(-1+\frac{2 \left(w^2+\kappa ^2\right) }{\left(\nu ^2+\rho '^2\right)}-\frac{3 w^2 \kappa ^2}{\left(\nu ^2+\rho '^2\right)^2}\right)  \tilde{x}^2  \,.
\ea\ee
This is exactly the same as the fluctuated Lagrangian \eqref{ell2}, and also \eqref{tanfol}. Because the characteristic frequencies are invariant under the T-duality transformation, the fluctuated Lagrangian of the long string limit in the reduced theory produces the correct frequencies. 

Recalling that the T-dual of the tanh model is the coth model in the reduced theory, we can expect the fluctuation Lagrangian of the coth model exactly agrees with that of the Nambu action. 
We are now interested in the nontrivial sector described by the complex sinh-Gordon model. Hence, for our present purpose, 
it is enough to perturb the Lagrangian of the coth model rather than the deformed gWZW model.
Plugging the classical solution \eqref{solcoth} into the fluctuation Lagrangian \eqref{quadcoth} and rescaling $\chi_{\!_A}$ such that its kinetic term has unit coefficient, 
we obtain the following Lagrangian, 

\be\ba{c}
\mathcal{L}_{{\rm coth}(2)}=\partial_+\delta \phi\partial_-\delta \phi-
 \left(\nu ^2+2 \rho'^2+\frac{2 w^2 \kappa ^2}{\nu ^2+\rho'^2}+ \frac{w^2 \kappa ^2\nu ^2}{\left(\nu ^2+\rho'^2\right)^2}\right) (\delta \phi)^2 \hspace{30pt} \\ \hspace{42pt}
+\partial_-\delta\chi_{\!_A} \partial_+\delta\chi_{\!_A} - \left(\nu ^2 +\frac{2 \left(w^2-\nu ^2\right) \left(\kappa ^2-\nu ^2\right)}{\rho'^2}
-\frac{2 w^2 \kappa ^2}{\nu ^2+\rho'^2}+\frac{w^2 \kappa ^2\nu ^2}{\left(\nu ^2+\rho'^2\right)^2}\right) (\delta \chi_{\!_A}^2) \\
-\frac{4 w \kappa  \nu ^3  \rho ''}{\rho' \left( \nu ^2+\rho'^2 \right)^2}\delta \chi_{\! _A} \delta \phi
+\frac{2 w \kappa  \nu  \left(\partial_-\delta\chi_{\!_A}-\partial_+\delta\chi_{\!_A} \right)}{\nu ^2+\rho'^2}\delta \phi \,.\hspace{62pt}
\ea\ee
In the Nambu action in the original string theory this Lagrangian is obtained if two fluctuations are introduced in the following way, 
\be \ba{c}
	t=\kappa \tau +N_1^1 z_1+N_2^1 z_2\,,~~~~\rho =\rho (\sigma) +N_1^2 z_1+N_2^2 z_2\,,\\
	\phi=\omega \tau +N_1^3 z_1+N_2^3 z_2\,,~~~~~~\varphi =\nu \tau +N_1^4 z_1+N_2^4 z_2\,,
\ea\ee 
where $N_1^i$ and $N_2^i$ are defined as
\be\ba{rl}
	& N_1=\left( \frac{w \tanh \rho(\sigma)}{\sqrt{\rho'(\sigma)^2 +\nu^2}} ,0,\frac{w \coth \rho(\sigma)}{\sqrt{\rho'(\sigma)^2 +\nu^2}} ,0 \right)\,,\\
	& N_2=\left( \frac{\kappa \nu}{\rho'(\sigma) \sqrt{\rho'(\sigma)^2+\nu^2}} ,0,  \frac{w \nu}{\rho'(\sigma) \sqrt{\rho'(\sigma)^2+\nu^2}} , \frac{\sqrt{\rho'(\sigma)^2+\nu^2}}{\rho'(\sigma) }  \right)\,.
\ea\ee
Substituting these into the Nambu action \eqref{nago}, one finds that $z_1$ corresponds to $\delta \phi$ and $z_2$ does $\delta \chi_{\!_A}$. 
Alternatively one can reproduce the same Lagrangian by applying $O(2)$ rotation to the two fields, $\tilde{\phi}$ and $\tilde{\varphi}$, 
in the Lagrangian \eqref{ngper}. 
Hence it turns out that the perturbation in the coth model corresponds to the perturbation of the Nambu action in the specific gauge. 

In the original theory or in the coth model, the Lagrangian possesses the term proportional to $1/\rho'^2$, while such a term does not appear 
in the tanh model. This can be understood by looking at the T-duality transformation in the reduced theory, \eqref{tdual}. For the present $\phi_{\!_A}$ in \eqref{soltanh}, we have $\tanh^2\phi_{\!_A}=\rho'^2/(\nu^2+\rho'^2)$ and the T-duality transformation is singular at turning points of the folded string. 
Because the partition function is invariant under the T-duality transformation, the partition function of the reduced theory is the same as that of the original string theory for the $(S,J)$ folded string, which supports our conjecture on the quantum partition function \eqref{eqpa}. 

Due to this direct relation between the coth model and the Nambu action one might think that it would be better to use the embedding of the coth model, \eqref{gcoth}, \eqref{gacot} rather than tanh model, \eqref{g0ta}, \eqref{gatan},  when comparing the fluctuations. 
However, as mentioned above, \eqref{gcoth} is a bad $H \x H$ gauge fixing for the perturbation. 
Hence we will basically continue to use the tanh model. 

In this section we have shown that the relation of the perturbations in the coth model, the tanh model and the Nambu action in the original string theory. 
Their fluctuation Lagrangians are seemingly different and the difference in the reduced theory follows form the choices of the complex sinh-Gordon model, or in other words, the choices of the $H\x H$ gauge.  
However, if a classical string solution in a subsector is described by the sin(sinh)-Gordon model, or equivalently, if the subgroup $H$ for the subsector is trivial, this freedom of choice does not exist, and so, the fluctuation Lagrangian of the reduced theory should always match that found by perturbing the Nambu action.
This is the case for the bosonic string theory in ${\rm R} \x {\rm S}^2$, which is studied in appendix \ref{apppuls}

\renewcommand{\theequation}{4.\arabic{equation}}
 \setcounter{equation}{0}

\section{Circular string\label{sjcs}}

In this section we shall discuss semiclassical quantization in the reduced theory for the $(S,J)$ circular string solution. Semiclassical computation in the original string theory was studied in \cite{art,ptt,gv,mikhcir}. 
It is expected that the perturbation in the reduced theory reproduces the result of \cite{ptt}. 

We shall derive the Lagrangian for the quadratic fluctuations by using the embedding of the tanh model, and evaluate their characteristic frequencies for the bosonic fluctuations in \ref{seccirbos} and for the fermionic fluctuations in \ref{seccirfer}. 
However a set of the characteristic frequencies in the reduced theory is not identical to the result in \cite{ptt}.
Then we shall show that the sum of the frequencies matches the original string theory result perturbatively in \ref{cirsumf}, and find a 2d Lorentz transformation such that each of the frequencies in the reduced theory is the same as the corresponding frequency in the original theory in \ref{seccir2d}. 
It may be considered that this happens because the $(S,J)$ circular string is the case of $\mu_+\ne\mu_-$ as a reflection of the existence of winding on a big circle of ${\rm S}^5$. As we will see in section \ref{mwfold} we observe the similar result for another example of $\mu_+\ne\mu_-$. 

We shall start the discussion with introducing the $(S,J)$ circular string solution in the embedding coordinates,  
\begin{equation} \label{sl2sol}\ba{c}
Y_0+iY_{3}=r_0  \,e^{i\kappa \tau}\,, 
~~~ Y_1+iY_2= r_1 \,e^{i {\rm w} \tau+ik\sigma}\,,  ~~~ X_1+iX_2=e^{i\omega \tau+im\sigma}\,,
\ea
\ee
where $r_0=\cosh \! \rho_0$ and $r_1=\sinh \! \rho_0$ with a constant radius $\rho_0$. 
$m$ and $k$ are integer winding numbers in the ${\rm AdS}_3$ subspace and the ${\rm S}^1$ subspace, respectively. 
This solution has three Cartan charges, $(E,S,J)=\sqrt{\lambda}(\mathcal{E},\mathcal{S},\mathcal{J})$, 
\be \mathcal{E}=r^2_0\kappa\,,~~~ \mathcal{S}=r^2_1w\,,~~~ \mathcal{J}=\omega \,.
\ee
The equations of motion read 
\be \label{eomcirs}
	{\rm w}^2=\kappa ^2+k^2\,,~~~ \omega^2 =\nu^2 + m^2\,,~~~ \nu^2=-\Lambda \,, ~~~ \kappa^2=\tilde{\Lambda}\,, 
\ee
whereas the conformal gauge constraints are written as
\be \label{sl2con} \ba{c}
	2\kappa \mathcal{E} -\kappa^2=2 \sqrt{k^2 +\kappa^2}+\mathcal{J}^2+m^2 \,, \\
	k\mathcal{S}+m\mathcal{J}=0 \,.
\ea\ee
They are supplemented by the identity $r_0^2-r_1^2=1$, which can be rewritten as
\be\label{sl2r0r1}
	\frac{\mathcal{E}}{\kappa}-\frac{\mathcal{S}}{\sqrt{k^2+\kappa^2}}=1\,.
\ee
The relations \eqref{eomcirs}, \eqref{sl2con} and \eqref{sl2r0r1} show that only three of these parameters are independent. 
When calculating quantum fluctuations it is convenient to use $\kappa$, $k$, $r_1$.\footnote{
Because $k$ can be absorbed in $\cal{S}$, it is possible to set $k$ to be $1$. Here we leave $k$ arbitrary in order to make it easier to compare our result with \cite{ptt}.}
From the latter three expressions, \eqref{sl2con} and \eqref{sl2r0r1}, we obtain the following relations, 
\be
\ba{c}
	\nu^2=\sqrt{\kappa^4-4 k^2\kappa^2 r_1^2(1+r^2_1) }\,,\\
	m^2=\ha \left( \kappa^2 - 2k^2 r_1^2- \nu^2 \right) \,.
\ea
\ee
whcih will be useful when we compute fluctuations. 

Now let us move on to the reduction of the $(S,J)$ circular string solution. 
The mass scale of the reduced theory $\mu$ is obtained through the observation of the stress tensor. 
For the $(S,J)$ circular string solution \eqref{sl2sol} we have 
\be
 T_{\pm \pm }^{\rm AdS}= -\left( \kappa ^2-2 k \left(k\pm\sqrt{k^2+\kappa ^2}\right) r_1^2 \right) \,,
\ee
which imply we should introduce $\mu_\pm^2=\sqrt{ \kappa ^2-2 k \left(k\pm\sqrt{k^2+\kappa ^2}\right) r_1^2}$. 
Because the closed string theory is defined on a cylinder rather than a plane, it is not allowed to set $\mu_+=\mu_-$ by using 2d Lorentz transformation.
Instead we proceed with a single $\mu$ defined by $\mu=\sqrt{\mu_+\mu_-}=\sqrt{\kappa } \left(\kappa ^2-4 k^2 r_1^2-4 k^2 r_1^4\right)^{1/4}$. 
Using this $\mu$, the relations in \eqref{red} for the tanh model give $\phi_{\!_A}$ and $\theta_{\!_A}$, 
\be \label{cirred} \ba{c}
 \phi_{\!_A} =\frac{1}{2} \log\left(\frac{\kappa -2 k r_1 \sqrt{1+r_1^2}}{\sqrt{\kappa ^2-4 k^2 r_1^2 \left(1+r_1^2\right)}}\right) \,, \\
 \theta_{\!_A} =A \left( \sqrt{k^2+\kappa ^2} \tau + k \sigma \right) \,,
\ea
\ee
where
\be\ba{c}
A=\frac{2  k^2   r_1^2 \left(1+r_1^2\right)}{\sqrt{\kappa } \left(\kappa ^2-4 k^2 r_1^2-4 k^2 r_1^4\right){}^{1/4} \left(\kappa -\sqrt{\kappa ^2-4 k^2 r_1^2-4 k^2 r_1^4}\right)}\,, 
\ea\ee
then the corrsponding classical solution in the deformed gWZW model is obtained by substituting these into \eqref{g0ta}, 
\be \label{g0sjcir}
g_0=
\left(
\begin{array}{cccc}
 0 & vB_+ & -vB_- & 0 \\
 - v^* B_+ & 0 & 0 &  v^* B_- \\
 vB_- & 0 & 0 & - v B_+ \\
 0 & -v^* B_- &  v^* B_+ & 0
\end{array}
\right) \,, \\
\ee
where
\be\label{vbsjcir}
\ba{c}
v=e^{iA \left( \sqrt{k^2+\kappa ^2} \tau + k \sigma \right) }\,,  \hspace{90pt} ~~~ 
\\\hspace{28pt}
B_\pm=\frac{1}{2}  \left(\frac{\sqrt{\kappa -2 k r_1 \sqrt{1+r_1^2}}}{\left(\kappa ^2-4 k^2 r_1^2 \left(1+r_1^2\right)\right){}^{1/4}}\pm\frac{\left(\kappa ^2-4 k^2 r_1^2 \left(1+r_1^2\right)\right){}^{1/4}}{\sqrt{\kappa -2 k r_1 \sqrt{1+r_1^2}}}\right) \,.
\ea
\ee
By solving the gauge field equations \eqref{redgaugef} we have the classical gauge fields, $A_{\pm 0}=\frac{i}{2}\, a_{\pm 0}R_2$ with 
\be
\ba{c}
a_{+0}=-\frac{\sqrt{\kappa } \left(k+\sqrt{k^2+\kappa ^2}\right)}{2 \left(\kappa ^2-4 k^2 r_1^2 \left(1+r_1^2\right)\right){}^{1/4}} \,, \\
a_{-0}=\frac{\left(k-\sqrt{k^2+\kappa ^2}\right) \left(\kappa ^2-4 k^2 r_1^2 \left(1+r_1^2\right)\right){}^{1/4}}{2 \sqrt{\kappa }} \,.
\ea
\ee
We find that $g_0^{-1}\partial_+g_0$ and $g_0^{-1}A_{+ 0}g_0$ are constant with these expressions. Hence this is a good starting point to discuss 
quantum fluctuations for the homogeneous string solution.

\subsection{Bosonic fluctuations in reduced theory\label{seccirbos}}

The ${\rm S}^5$ sector is rather simple; 
bosonic fluctuations in the ${\rm S}^5$ sector for a string solution in ${\rm AdS}_3 \x {\rm S}^1$ are 
massive fields with masses $\pm \mu$.  
In the present case we have $\mu^2=\mu_+\mu_-$, then we obtain characteristic frequencies for the four massive fields, 
\be
	\pm \sqrt{n^2+\mu^2}=\pm \sqrt{n^2+\sqrt{\kappa^4-4 k^2\kappa^2 r_1^2(1+r^2_1) }}\,, \label{sl2fres5}
\ee
which are exactly the same as the result in \cite{ptt}.

As done in the case of the folded string, we introduce the fluctuation fields by \eqref{etap}, \eqref{etao} and \eqref{dgau},\footnote{
Note that  $v$ in the expression \eqref{dgau} should be replaced by $v$ in \eqref{vbsjcir}. 
}
integrate out the diagonal parts of the gauge field fluctuations, 
and then, use the $H$ gauge freedom such that physical fluctuations decople from unphysical fluctuations. 
The physical part of the quadratic fluctuations is described by the Lagrangian containing $a_1$ and $a_2$,
\be \label{sl2l1}
\mathcal{L}_1=2\sum _{i=1,2}\left( \partial_+a_i \partial_- a_i -\kappa ^2a_i^2\right)\,,
\ee
and the Lagrangian containing $a_3$ and $a_4$, 
\be \label{sl2l2} \ba{c}
\mathcal{L}_2= 2\bigg[ \partial _- a_3 \partial _+ a_3
-2 \kappa  \left(\kappa -\sqrt{\kappa ^2-4 k^2 r_1^2 \left(1+r_1^2\right)}\right) a_3 ^2 +\partial _- a_4 \partial _+ a_4 
\\
~~~~~~~~~~~~+\frac{2 \left(\kappa ^2-4 k^2 r_1^2 \left(1+r_1^2\right)\right){}^{1/4}}{\sqrt{\kappa }} \left(\left(k+\sqrt{k^2+\kappa ^2}\right) \partial _- a_3-\left(k-\sqrt{k^2+\kappa ^2}\right) \partial _+ a_3\right) a_4
\bigg]
 \,.
\ea\ee
Now it is clear that $a_1$ and $a_2$ correspond to $\tilde{Y}_2$ in \cite{ptt}, whose frequencies are 
\be
	\pm \sqrt{n^2+\kappa^2} \label{sl2fres2} \,.
\ee
However the second Lagrangian $\mathcal{L}_2$ does not describe the remaining two fluctuations in ${\rm AdS}_5$ in \cite{ptt}. 
In fact, by substituting $e^{i(\Omega \tau-n \sigma)}$ into the equations of motion, 
one finds that the condition that the determinant of the mass matrix vanishes reads 
\be \label{freqeq} \ba{c}
(n^2-\Omega^2)^2-\frac{8 k  \sqrt{\left(k^2+\kappa ^2\right) \left(\kappa ^2-4 k^2 r_1^2(1+r_1^2) \right)}}{\kappa }n \Omega+ 
\frac{2  \left(\kappa ^3-(\kappa ^2 +2 k^2) \sqrt{\kappa ^2-4 k^2 r_1^2(1+r_1^2)}\right)}{\kappa }n^2 \\
-\frac{2  \left(\kappa ^3+(\kappa ^2 +2 k^2) \sqrt{\kappa ^2-4 k^2 r_1^2(1+r_1^2)}\right)}{\kappa }\Omega^2=0 \,,
\ea\ee 
which is different from the corresponding equations in the original theory (c.f., Eq. (4.15) in \cite{ptt}), 
\be \label{corboscir}
\left( \Omega^2-n^2 \right)^2+4 \Omega^2 \kappa ^2 r_1^2-4 \left(1+r_1^2\right) \left(\Omega \sqrt{k^2+\kappa ^2} + k n \right)^2 =0\,,
\ee
and consequently, the characteristic frequencies are different.\footnote{
Note that the equation \eqref{corboscir} is obtained from the equation (4.15) in \cite{ptt} by $\Omega \to -\Omega$. 
This difference originates from the fact we use the mode expansion $e^{i(\Omega \tau-n \sigma)}$ rather than $e^{i(\Omega \tau+n \sigma)}$. 
} 
Although the equation \eqref{freqeq} can not be solved for a genral case,  
it is possible to solve the equation \eqref{freqeq} approximately for large $\cal{J}$ with $u=\cal{S}/\cal{J}$, $k$ fixed.  
The four roots $\Omega_{I;n}$ are 
\be \label{sl2freads3} \ba{c}
	\Omega_{I=1,2;n}= \frac{-2 k n\pm \sqrt{n^2 \left(n^2+4 k^2 u (1+u)\right)}}{2 \mathcal{J}} + O\left( \frac{1}{\mathcal{J}^3} \right) \,, \\
	\Omega_{I=3,4;n}= \pm 2 \mathcal{J}\pm \frac{n^2\mp 2 k n + 2 k^2 (1+u)}{2 \mathcal{J}}+ O\left( \frac{1}{\mathcal{J}^3} \right) \,.
\ea\ee
For $n=0$ the equation \eqref{freqeq} can be solved exactly, 
\be \label{sl2zero}
\Omega_{I,0}=\{0,0, \Omega_0,-\Omega_0\}\,, ~~~\Omega_0 = \sqrt{\frac{2\kappa^3 +2(\kappa ^2 +2 k^2 ) \sqrt{\kappa ^2-4 k^2 r_1^2 \left(1+r_1^2\right)}}{\kappa }}\,.
\ee
These frequencies are totally different from those in \cite{ptt}.
Therefore, as far as the individual characteristic frequencies are concerned, the reduced theory does not reproduce the result of the original string theory even for the zero modes.

For consistency we shall discuss the perturbation in the coth model which should describe the nontrivial sector containing $a_3$ and $a_4$, \eqref{sl2l2}. 
Using the reduction \eqref{redcoth} and the fluctuation Lagrangian of the coth model \eqref{quadcoth} we obtain the following Lagrangian, 
\be \ba{c}\mathcal{L}_{\rm coth}=
\partial_+ \delta \phi \partial_- \delta \phi -2 \kappa   \left(\kappa +\sqrt{\kappa ^2-4 k^2 r_1^2 \left(1+r_1^2\right)}\right) \delta \phi^2 +\frac{\kappa +\sqrt{\kappa ^2-4 k^2 r_1^2 \left(1+r_1^2\right)}  }{\kappa -\sqrt{\kappa ^2-4 k^2 r_1^2 \left(1+r_1^2\right)}} \partial_- \delta \chi_{\!_A}  \partial_+ \delta \chi_{\!_A} \\
+16 k^2 r_1^2 \left(1+r_1^2\right) \left(\kappa ^2-4 k^2 r_1^2 (1+r_1^2) \right){}^{1/4} \left(\kappa -2 k r_1 \sqrt{1+r_1^2}\right) \hspace{30pt} \\
 \hspace{30pt} \times  \,
\frac{ \left(\kappa -2 k r_1 \sqrt{1+r_1^2}+\sqrt{\kappa ^2-4 k^2 r_1^2(1+r_1^2) }\right) \left(\left(k+\sqrt{k^2+\kappa ^2}\right)  \partial_- \delta \chi_{\!_A} +\left(k-\sqrt{k^2+\kappa ^2}\right) \partial_+ \delta \chi_{\!_A} \right)}{\sqrt{\kappa } \left(\kappa -2 k r_1 \sqrt{1+r_1^2}-\sqrt{\kappa ^2-4 k^2 r_1^2 (1+r_1^2)}\right){}^3 \left(\kappa +\sqrt{\kappa ^2-4 k^2 r_1^2 (1+r_1^2) }\right)} \delta \phi \,,
\ea\ee
which yields the same equation for characteristic frequencies as \eqref{freqeq}. 
Hence the bosonic frequencies for this subsector is not model-dependent. 

Because the other six bosonic frequencies match the corresponding six frequencies of the original theory, the discrepancy in these two frequencies seems to be a serious problem. 
However, in the following subsections, we will show that the total sum of the characteristic frequencies including the fermionic contributions is the same as that of the original string theory. 
Hence the discrepancy here is not indeed a problem.

\subsection{Fermionic fluctuations in reduced theory\label{seccirfer}}
We define component fields of the fermionic fluctuations as in the folded string case \eqref{ferRL1} - \eqref{ferRL5}.\footnote{
Here we should use $v$ in \eqref{vbsjcir} for $v$ in the expression of $\delta \Psi_{_L}$.}
Then the fermionic part of the quadratic fluctuations takes the form, 
\be \label{sl2fer2} \ba{l}
	{\cal L}_{f}=2 \Big[ \sum _{i=1}^8\left( \alpha_i \dm \alpha_i +\beta_i \dm \beta_i \right) + \frac{\left(k-\sqrt{k^2+\kappa ^2}\right) \left(\kappa ^2-4 k^2 r_1^2(1+r_1^2)\right){}^{1/4}}{\sqrt{\kappa }}\left(-\alpha_1 \alpha_2 - \alpha_3 \alpha_4 -\alpha_5 \alpha_6 +\alpha_7 \alpha_8  \right)\nonumber \\ 
	\hs{30pt}
	 + \frac{\left(k+\sqrt{k^2+\kappa ^2} \right) \left(\kappa ^2-4 k^2 r_1^2(1+r_1^2)\right){}^{1/4}}{\sqrt{\kappa }} \left(- \beta_1 \beta_2-\beta_3 \beta_4-\beta_5 \beta_6+\beta_7 \beta_8 \right)\nonumber \\ 
	\hs{30pt}
	+ \sqrt{\kappa } \left(\sqrt{\kappa -2 k r_1 \sqrt{1+r_1^2}}+\sqrt{\kappa +2 k r_1 \sqrt{1+r_1^2}}\right) \nonumber \\ 
	\hs{50pt} \x \left( -\alpha_3\beta_1  -\alpha_1\beta_3  +\alpha_7\beta_5  -\alpha_5\beta_7  +\alpha_4\beta_2  +\alpha_2\beta_4 + \alpha_8\beta_6 - \alpha_6\beta_8  \right)  \Big] \,.
\ea\ee
As expected the coefficients of each term in the Lagrangian is totally constant. 
Then we can evaluate frequencies in a straightforward way, 
\be
\pm\sqrt{\left( n \pm c\right)^2+ a^2}\pm d\,,
\ee
where 
\be
\ba{rl}
&a^2=\frac{\kappa}{2}   \left(\kappa +\sqrt{\kappa ^2-4 k^2 r_1^2(1+r_1^2)}\right)\,, \\
&c=\frac{k \left(\kappa ^2-4 k^2 r_1^2 \left(1+r_1^2\right)\right){}^{1/4}}{2 \sqrt{\kappa }} \,, \hspace{51pt}\\
&d=\frac{\sqrt{  k^2+\kappa ^2 } \left(\kappa ^2-4 k^2 r_1^2 \left(1+r_1^2\right)\right){}^{1/4}}{2 \sqrt{\kappa} } \,. \hspace{7pt}
\ea
\ee
While $a$ agrees with $a$ in \cite{ptt}, the other two do not match. 
 
Despite of the discrepancies in the frequencies, one might still expect that the sum of all the frequencies match, which 
can provide a nontrivial support that the partition functions in the two theories are equivalent in the case of the $(S,J)$ circular string at one-loop level.  
To show this we shall evaluate the sum of the frequencies perturbatively in $\cal{J}$ in the next subsection.

\subsection{Sum of frequencies\label{cirsumf}}

The procedure of calculating sum of the characteristic frequencies 
directly follows from the computation of the one-loop energy correction discussed in \cite{ptt}.  
If given a set of $N$ fluctuations, we obtain $2N$ roots $\Omega_{I;n}$ ($I=1,\dots, 2N$) by solving the conditions that the determinant 
of the corresponding $N \x N$ mass matrix vanishes.  
The zero modes appear in pairs, $\Omega_{I;0}={\pm \Omega_{p;0}}$ ($p=1,\dots, N$), and the non-zero modes can be paired 
by the condition $\Omega_{I;n}=-\Omega_{I;n}$.
Then the frequencies should be summed up as
\be
	\sum_{p=1}^{N} \hat  \Omega_{p;0}
   +\sum_{n=1}^\infty\;\sum_{I=1}^{2N} \hat \Omega_{I;n} \,,
\ee
with  
\be
\hat  \Omega_{p;0}
 =   {\rm sign}(C_{p}) \Omega_{p;0}\, , \ \ \ \ \ \ \ \
\hat \Omega_{I;n} =  {\rm sign}(C_{I}^{(n)})  \Omega_{I;n} \, , \ee
 \begin{equation}
C_{p}=\frac{1}{2m_{11}(\Omega_{p;0})\Omega_{p;0} \prod_{q\neq
p}(\Omega_{p;0}^{2}-\Omega _{q;0}^{2})}\,,\quad
C_{I}^{(n)}=\frac{1}{m_{11}(\Omega_{I;n})\prod _{J\neq
I}(\Omega_{I;n}-\Omega_{J;n})}\, ,  \label{signs}
\end{equation}
where $m_{11}$ is a minor of the mass matrix, i.e., the determinant of the
matrix obtained from the mass matrix by removing the first row and first
column.
Note that the fermionic frequencies contribute to the partition function negatively.
For the $(S,J)$ string we find the structure of ${\rm sign}C_I^{(n)}$ becomes simple because half of the frequencies are positive and half of the frequencies are negative. 

The zero modes contribution is
 \be
  4 \nu  + 2\kappa  +  \Omega_0  - 8  \sqrt{c^2+a^2}  \, , \la{zero}
 \ee
 and nonzero modes are 
 \bea \la{nonz} 2
 \sum_{n=1}^\infty
 \bigg[ 4 \sqrt{n^2 + \nu^2}+ 2 \sqrt{n^2 + \k^2}
 + \ha  \sum^{4}_{I=1} {\rm sign}(C_{I}^{(n)}) \Omega_{I;n} 
  - \ 4 \big( \sqrt{  ( n + c)^2 + a^2}
   + \sqrt{ (n - c)^2 + a^2 }\ \big) \bigg] \, ,  \eea
where $\Omega_0$ is shown in \eqref{sl2zero} and $\Omega_{I;n}$ is expanded in terms of large $\mathcal{J}$ in \eqref{sl2freads3}.
Using the formula for the large $\mathcal{J}$ expansion of $\kappa$, 
\be
\kappa =\mathcal{J}+\frac{k^2 u (2+u)}{2 \mathcal{J}}-\frac{k^4 u \left(4+12 u+8 u^2+u^3\right)}{8 \mathcal{J}^3}+\cdots \,,
\ee
where $u=\cal{S}/\cal{J}$, 
we find the zero mode part, 
\be
-\frac{k^2 u(1+ u)}{\mathcal{J}}+O\left(\frac{1}{\mathcal{J}^3}\right) \,.
\ee
On the other hand, the contribution from the non-zero modes is expanded as
\be
-\sum_{n=1}^\infty\frac{n^2+2k^2 u (1+u)- n \sqrt{n^2+4 k^2 u (1+u)}}{2\mathcal{J}} +O\left(\frac{1}{\mathcal{J}^3}\right) \,.
\ee
These two results are consistent with \cite{ptt}. 

It is still mysterious that the characteristic frequencies of each fluctuation do not match. 
In \cite{ptt} the expansions of the Landau-Lifshitz Lagrangian is also discussed as a useful tool for extracting the part of the fluctuation frequencies in the string theory, $\Omega_{I=1,2}$. Although each frequency in the Landau-Lifshitz model is different from the corresponding one in the string theory, the sum of the two frequencies agrees with the string theory result. 
So one might expect that the sum of the frequencies of $a_3$ and $a_4$ would agree with that of the two fluctuations in the original string theory 
in the large $\mathcal{J}$ expansion. 
However this does not happen. Actually, for the nonzero modes of $a_3$ and $a_4$, we have  
\be
 \ha  \sum^{4}_{I=1} {\rm sign}(C_{I}^{(n)}) \Omega_{I;n} =2 \mathcal{J}+\frac{2 k^2(1+ u)+n\left( n+ \sqrt{n^2+4 k^2 u (1+u)}\right)}{2 \mathcal{J}} +O\left(\frac{1}{\mathcal{J}^3}\right)\,, \label{sum2red}
\ee
On the other hand, the corresponding sum in the original string theory is   
\be
 2 \mathcal{J}+\frac{2 k^2 \left(1+3 u+u^2\right)+n \left(n+\sqrt{n^2+4 k^2 u (1+u)}\right)}{2 \mathcal{J}}+O\left(\frac{1}{\mathcal{J}^3}\right)\,,
\ee
which is different from \eqref{sum2red} and the discrepancy is 
\be
\frac{k^2 u (2+u)}{\mathcal{J}} \,.
\ee
This cancels with the discrepancy in the fermionic sector, and then the total sums of the frequencies are the same.

\subsection{2d Lorentz boost\label{seccir2d}}

Generally we can evaluate the characteristic frequencies and confirm the agreement of the total sum of the frequencies order by order in the large $\mathcal{J}$.
However it is technically hard to continue the calculation to higher orders in $1/\mathcal{J}$. 
Here we shall find a $2d$ Lorentz transformation on the worldsheet, which does not change the total sum of the bosonic and fermionic frequencies, 
but can change some of the frequencies.

Because the equations for the two fluctuations $a_1$, $a_2$ in the ${\rm AdS}_5$ sector and all of the four fluctuations in the ${\rm S}^5$ sector do not contain the first derivative term, their frequencies are obviously invariant under the Lorentz boost on the worldsheet.  
Hence one can expect a certain 2d Lorentz boost allows us to modify the equations for the fermionic fluctuations and the other two bosonic fluctuations such that they yield the frequencies found in \cite{ptt}.
Let us introduce the 2d Lorentz boost by 
\be
	\tau \to p_1 \tau + q_1 \sigma \,, ~~~\sigma \to q_1 \tau + p_1 \sigma \,, ~~~{\rm with}~~~p_1^2-q_1^2=1\,.
\ee
In order to set the equations for frequencies of the two bosonic fluctuations $a_3$ and $a_4$ to be the same as the corresponding equation in the original theory \eqref{corboscir}, we should choose
\be \label{booscir} \ba{c}
 p_1=\sqrt{\frac{\kappa ^2-2 k^2 r_1^2+\kappa  \sqrt{\kappa ^2-4 k^2 r_1^2 \left(1+r_1^2\right)}}{2 \kappa  \sqrt{\kappa ^2-4 k^2 r_1^2 \left(1+r_1^2\right)}}} \,, \\
 q_1=-\sqrt{\frac{\kappa ^2-2 k^2 r_1^2-\kappa  \sqrt{\kappa ^2-4 k^2 r_1^2 \left(1+r_1^2\right)}}{2 \kappa  \sqrt{\kappa ^2-4 k^2 r_1^2 \left(1+r_1^2\right)}}} \,.
\ea\ee 
By this 2d boost the fermionic frequencies are changed into 
\be
\pm\sqrt{\left( n \pm c\right)^2+ a^2}\pm d\,,
\ee
where 
\be
\ba{rl}
&a^2=\frac{ \left(\kappa^2 +\kappa \sqrt{\kappa ^2-4 k^2 r_1^2 \left(1+r_1^2\right)}\right)}{2}\,,  \\
&c=\frac{\sqrt{\kappa ^2+2 k^2 \left(1+r_1^2\right)-\kappa  \sqrt{\kappa ^2-4 k^2 r_1^2 \left(1+r_1^2\right)}}}{2 \sqrt{2}} \,, \\
&d=\frac{\sqrt{\kappa ^2+2 k^2 \left(1+r_1^2\right)+\kappa  \sqrt{\kappa ^2-4 k^2 r_1^2 \left(1+r_1^2\right)}}}{2 \sqrt{2}} \,, 
\ea
\ee
which are exactly the same as \cite{ptt}. 
Therefore all of the fluctuations in the reduced theory agree with the result of \cite{ptt} in this frame. 
Because the sum of the frequencies should be invariant under the 2d boost, our result implies that the sum of the frequencies in the reduced theory, \eqref{zero} and \eqref{nonz}, recovers the result of the original string theory calculation to all orders in $1/\mathcal{J}$. 
Then the quantum equivalence of the partition functions \eqref{eqpa} is proven for the $(S,J)$ circular string at the one-loop level. 

Let us discuss the meaning of this $2d$ Lorentz boost. 
Because the form of the classical solution \eqref{sl2sol} is not $2d$-Lorentz invariant, the fluctuation Lagrangians \eqref{sl2l2} and \eqref{sl2fer2} are not either, and consequently, each fluctuation frequency may change by the $2d$ Lorentz transformation. 
If we apply the Lorentz transformation \eqref{booscir} to the classical solution \eqref{sl2sol}, the solution becomes 
\be \label{cirmodi}
\ba{c}
Y_0+iY_{3}=r_0  \,e^{i\kappa p_1 \tau +i \k q_1 \sigma }\,, 
~~~ Y_1+iY_2= r_1 \,e^{i ({\rm w} p_1+kq_1) \tau+i(k p_1+{\rm w} q_1)\sigma}\,, \\ X_1+iX_2=e^{i(\omega p_1 +m q_1)\tau+i(m p_1+\omega q_1)\sigma}\,.
\ea
\ee
With this classical solution the stress tensor takes the following form, 
\be
 T_{\pm \pm }^{\rm AdS}=-\kappa  \sqrt{\kappa ^2-4 k^2 r_1^2 \left(1+r_1^2\right)} \,,
\ee
which implies $\mu_+=\mu_-=\sqrt{\kappa } \left( \kappa ^2-4 k^2 r_1^2 \left(1+r_1^2\right)\right)^{1/4}$. 
Hence it turns out that the $2d$ Lorentz boost we applied to both the bosonic fluctuations and the fermionic fluctuations is the same as the one setting $\mu_+=\mu-$.  
However, it is still mysterious why the mixing of the bosonic and fermionic frequencies occurs by the Pohlmeyer reduction and why it is necessary to set $\mu_+=\mu_-$ for the agreement of the frequencies of the individual fluctuations.


\renewcommand{\theequation}{5.\arabic{equation}}
 \setcounter{equation}{0}

 \section{Spiky string\label{sjspik}}

We shall discuss semiclassical expansion around the $(S,J)$ spiky string solution in ${\rm AdS}_3 \x {\rm S}^1$ by using the embedding of the tanh model into the deformed gWZW model for the bosonic sector in \ref{secspikbos} and for the fermionic sector in \ref{secspikfer}. 
Classical aspects of the spiky string solutions were studied in the bosonic string theory \cite{krusp, iktt} and in the Pohlmeyer-reduced form \cite{jevjin2}. 
In \cite{hit} and in the earlier sections of this paper we have seen that the semiclassical computation in the reduced theory perfectly recovers the one-loop corrections to the string partition function, and moreover, it has a huge advantage as the reduced theory has simple structures of both the bosonic and fermionic fluctuations after properly fixing the $H$ gauge.    
Hence we expect that our result will agree with the string theory side, and the calculation here will be much simpler than the standard worldsheet approach in the conformal gauge string theory.

First we shall review the ($S,J$) spiky string solution in ${\rm AdS}_3 \x {\rm S}^1$ found the paper \cite{iktt}. 
The solution is expressed in terms of the embedding coordinates,\footnote{
Here we have assumed $r_0$ and $r_1$ have the $u(=\alpha \sigma +\beta \tau)$ dependence whereas they were constants in the case of the circular string \eqref{sl2sol}. 
} 
\be\ba{c}
	Y_0+iY_3=r_0(u)\, e^{iw_0 \tau +i\varphi_0(u)}\,,~~~Y_1+iY_2=r_1(u)\, e^{iw_1 \tau +i\varphi_1(u)}\,,~~~X_1+iX_2=e^{i\psi (u)}\,,
\ea\ee
with
\be
u=\alpha \sigma +\beta \tau \,,~~~~~r_0^2-r^2_1=1\,. 
\ee
Here $w_0$ and $w_1$ are real constants.  
The ${\rm S}^1$ part is explicitly written as
\be
	\psi = \nu \tau +\frac{D-\beta \nu}{\beta^2-\alpha^2}u \,, 
\ee
while $\varphi_0$ and $\varphi_1$ are expressed in differential form, 
\be\ba{c}
	\varphi'_0 = -\frac{1}{\beta^2-\alpha^2}\left( \frac{C_0}{r_0^2}+w_0 \beta \right) \,, \\
	\varphi'_1 = \frac{1}{\beta^2-\alpha^2}\left( \frac{C_1}{r_1^2}+w_1 \beta \right) \,, 
\ea\ee
where $\nu$, $D$, $C_0$ and $C_1$ are real constants. 
The Virasoro constraints read 
\be \label{spvir}\ba{c}
-r_0^{'2}(\beta^2-\alpha^2)-\frac{C_0^2}{r_0^2(\beta^2-\alpha^2)}-
\frac{r_0^2\alpha^2w_0^2}{\beta^2-\alpha^2} +r_1^{'2}(\beta^2-\alpha^2)+\frac{C_1^2}{r_1^2(\beta^2-\alpha^2)}+ \frac{r_1^2 \alpha^2
w_1^2}{\beta^2-\alpha^2}+\frac{D^2}{\beta^2-\alpha^2}+\frac{\alpha^2\nu^2}{\beta^2-\alpha^2}=0\,,
\ea\ee
and
\be
w_0 C_0+w_1 C_1+D\nu=0 \,. \label{spvir2}
\ee
The first constraint is nothing but the condition that the Hamiltonian of this system should vanish. 
Using $r_0^2-r_1^2=1$ we rewrite \eqref{spvir} into an equation for $r_1$, 
\begin{equation}
(\beta^2-\alpha^2)^2
r_1^{'2}=(1+r_1^2)\bigg(\frac{C_0^2}{1+r_1^2}+\alpha^2 w_0^2
(1+r_1^2)-\frac{C_1^2}{r_1^2}-\alpha^2 w_1^2 r_1^2 -D^2-\alpha^2
\nu^2\bigg) \,. \label{r1}
\end{equation}
Here it is necessary to assume $w_0^2<w_1^2$ such that the string does not attach the boundary.

The spiky string solution with $n$ spikes consists of $2n$ arcs, each of which should possess two turning points ($r'_1=0$) at some finite values of $r_1$. 
Let us introduce a new radial variable $v(u)$ by 
\be
	v=\frac{1}{1+2 r_1^2}=\frac{1}{\cosh 2 \rho } \,, \label{rtovtorho}
\ee
where $\rho$ is the radial coordinate in the global coordinate system of ${\rm AdS}_3$. 
Assume that $v'$ vanishes at $v=v_1, v_2, v_3$ with $ v_1\leq0\leq v_2\leq v_3\leq 1$, then the equation for $r_1$ \eqref{r1} is rewritten as 
\be
	v'=\frac{\sqrt{2vP(v)}}{\alpha^2-\beta^2}\,, ~~~~~
	P(v)=\frac{ w_1^2-w_0^2 }{v_1v_2v_3}\left(v-v_1 \right)\left(v-v_2 \right)\left(v-v_3 \right)\,.
\ee
The constant $v_1$ is not arbitrary but is a function of $v_2$ and $v_3$, 
\be
	v_1=-\frac{v_2v_3}{v_2+v_3+v_2v_3\frac{w_1^2+w_0^2-2(\nu^2 +D)}{w_0^2-w_1^2}}\,.
\ee
$C_1$ and $C_2$ are expressed in terms of $v_1$, $v_2$ and $v_3$
\be\ba{c}
	C_0^2= \frac{w_0^2-w_1^2}{8} \frac{\left(1+v_1 \right)\left(1+v_2 \right)\left(1+v_3 \right)}{v_1v_2v_3}\,, \\
	C_1^2=\frac{w_0^2-w_1^2}{8} \frac{\left(1-v_1 \right)\left(1-v_2 \right)\left(1-v_3 \right)}{v_1v_2v_3}\,.
\ea\ee
For our assumption $ v_1\leq0\leq v_2\leq v_3\leq 1$ and $w_0^2<w_1^2$, we have $C_0^2\geq 0$ and $C_1^2\geq 0$, then 
our choice of the roots is consistent.  
Depending on further conditions for $v_2$ and $v_3$, the solution has two possible regimes; the spike is at the minimum $r_1$ in one regime and at the maximum 
$r_1$ in the other regime. 

Next let us discuss the reduction of the ($S,J$) spiky string solution. 
The mass scale $\mu$ in the reduced theory is read out from ${\rm AdS}_3$ part of the stress tensor, $T_{\pm \pm}^{\rm AdS}$, in the original string theory.  
For the $(S,J)$ spiky string solution, due to its complication in the AdS sector, it is convenient to calculate $T_{\pm \pm}^{\rm AdS}$ from the ${\rm S}^1$ part of the stress tensor $T_{\pm \pm}^{\rm S}$ by using the Virasoro constraint, $T_{\pm \pm}^{\rm AdS}+T_{\pm \pm}^{\rm S}=0$. Then we have   
\be\ba{c}
	T_{\pm \pm}^{\rm AdS}=-T_{\pm}^{\rm S}=-\left(\frac{\alpha  \nu \mp D}{\alpha \mp \beta } \right)^2\,,
\ea\ee
which show this is also the case of $T_{++}^{\rm AdS} \ne T_{--}^{\rm AdS}$. So we should introduce $\mu_\pm$ by $T_{\pm \pm}=-\mu_\pm^2$,  
and define the mass scale $\mu $ as 
\be\label{spimu}
	\mu=\sqrt{\mu_+\mu_-}=\sqrt{\frac{\alpha ^2 \nu ^2-D^2}{\alpha ^2-\beta ^2}} \,.
\ee
By this definition of $\mu$ we have implicitly assumed $\alpha ^2 \nu ^2 > D^2$ and $\alpha ^2>\beta ^2$ (or $\alpha ^2 \nu ^2 < D^2$ and $\alpha ^2<\beta ^2$) 
so that $\mu$ is real. 
So we will hereafter work on the reduction in these parameter regions. 
Although the corresponding solution in the reduced theory takes a different form for other parameter regions, one can show that the resulting fluctuation Lagrangian is independent of the choice of the parameters. 

Following the standard procedure of the Pohlmeyer reduction for the tanh model, \eqref{red}, we obtain the sinh-Gordon angle $\phi_{\!_A}$,\footnote{For notational simplicity we will use $r_1$ rather than $v$ which is used in the original paper \cite{iktt}. 
Although expressing the classical solution in terms of $r_1'$ makes it easy to understand the behavior of the fluctuations at spikes, we avoid to employ $r_1'$ for the same reason.}
\be \label{phispik} \ba{c}
\phi_{\!_A}=
\frac{1}{2} \log\left[\frac{M_2+2 \alpha  \sqrt{M_2 M_3} +\alpha ^2 M_3 }{\alpha ^2 \nu ^2-D^2}\right] \,,
\ea\ee
and $\theta_{\!_A}$, 
\be \label{pmtheta} \ba{c}
\partial _\pm\theta_{\!_A} =\alpha ^2 \sqrt{\frac{\alpha ^2-\beta ^2}{\alpha ^2 \nu ^2-D^2}} \frac{ (C_0^2+C_1^2) w_0 w_1 
+  C_0 C_1 \left(w_0^2+w_1^2\right)\pm C_0 w_1 \alpha \left(M_1-w_0^2\right)  
 \pm C_1 w_0 \alpha \left(M_1-w_1^2\right)  +w_0 w_1 \left( \alpha ^2 M_3 -D^2 \right)}{ (\alpha \mp \beta )^2M_2 }\,,
\ea\ee
where we have introduced $M_i~(i=1,2,3)$ as functions of $r_1$, 
\be
M_1=  w_0^2+ \left(w_0^2-w_1^2\right) r_1^2 \,,
\ee
and
\be\ba{c} 
M_2= M_1 \alpha ^2-D^2 \,, ~~~~
M_3= M_1-\nu ^2 \,.
\ea\ee
Hereafter let us consider the case of positive $M_1$.\footnote{
This assumption is related to a condition for the existence of the solution of the equation \eqref{r1} for $D =\nu = 0$. 
Naively the equation \eqref{r1} has a solution if its right hand side is positive. 
A sufficient condition for this is exactly the same as $M_1 \geq 0$ because we have $C_0^2\geq C_1^2$.

If we consider the other case where $M_1$ is negative, the $D,\nu \to 0$ limit in the fluctuation Lagrangians yields a wrong answer. 
This problem is solved by the following prescription. 
Generally the reduction equation for $\phi_{\!_A}$ in \eqref{red} has four solutions, two of which are real. 
Employing the other real branch for $\phi_{\!_A}$, 
\be
 \frac{1}{2} \log\left[\frac{M_2-2 \alpha  \sqrt{M_2 M_3} +\alpha ^2 M_3 }{\alpha ^2 \nu ^2-D^2}\right] \,,
\ee 
one can check the $D,\nu \to0$ limit becomes well-defined for $M_1<0$ at the level of the fluctuation Lagrangians.  
This observation implies that we should choose $\phi_{\!_A}$ appropriately depending on the parameters, $\alpha$ and $w_0$. 

Another reason we have assumed $M_1 \geq 0$ is that the solution has a smooth limit to the folded string in ${\rm AdS}_3$ under this assumption. 
In fact the folded string limit corresponds to $M_1 \to \rho'^2$ which is positive-definite.
} 
Because the radial coordinate $r_1$ is a function of $u=\alpha \tau +\beta \sigma$, $\theta_{\!_A}$ can be expressed in an integral form. 
Making the T-duality transformation does not help in the present case; $\partial _\pm \theta_{\!_A}$ are not constant also in the coth model. 
However, this is not a serious problem as far as quadratic fluctuations are concerned, since $\theta_{\!_A}$ appears only as $\partial _\pm \theta_{\!_A}$ in the fluctuation Lagrangian. 
 
If we substitute \eqref{phispik} and \eqref{pmtheta} into \eqref{g0ta} we find that the corresponding classical solution in the gWZW model takes the following form, 
\be\label{g0spik}
g_0=
\left(
\begin{array}{cccc}
 0 & vP_1 & -vP_2 & 0 \\
 - v^* P_1 & 0 & 0 &  v^* P_2 \\
 vP_2 & 0 & 0 & - v P_1 \\
 0 & -v^* P_2 &  v^* P_1 & 0
\end{array}
\right) \,. \\
\ee 
$v$ is given by
\be \label{thespik}
v=e^{i\theta_{\!_A} }\,, 
\ee
where $\theta_{\!_A}$ solves the differential equations in \eqref{pmtheta}. 
$P_1$ and $P_2$ are written as 
\be\label{b1b2spik}\ba{c}
P_1=\frac{M_2+\alpha \sqrt{M_2 M_3}  }{\sqrt{\left(M_2+  \alpha ^2 M_3 +2 \alpha \sqrt{M_2 M_3}  \right) \left(\alpha ^2 \nu ^2-D^2\right)}}\\
P_2=\frac{ \alpha^2 M_3 +\alpha\sqrt{M_2 M_3} }{\sqrt{\left(M_2+\alpha ^2 M_3 +2 \alpha \sqrt{M_2 M_3}\right) \left(\alpha ^2 \nu ^2-D^2\right)}}
\ea\ee
The classical gauge field equations \eqref{redgaugef} are solved by $A_{\pm 0}=\frac{i}{2}\, a_{\pm 0}R_2$ with 
\be
a_{\pm 0}=-\frac{\left(M_2 \pm \alpha ^2 M_3 \right) \partial_\pm \theta_{\!_A}}{ M_2} \,,
\ee
where $\partial_\pm \theta_{\!_A}$ are given by \eqref{pmtheta}.

At the level of the classical solution \eqref{g0spik} we can not take the $D,\nu \to 0$ limit in which the string is not stretching in ${\rm S}^5$. 
It is  because $P_{1,2}$ in \eqref{b1b2spik} diverge in this case due to the factor of $\sqrt{\alpha ^2 \nu ^2-D^2}$ in their denominators. 
We will show the $D,\nu \to 0$ limit can be taken once we derive the Lagrangian for quadratic fluctuations, which is the same situation as the $(S,J)$ folded string case.

  \subsection{Bosonic fluctuations in reduced theory\label{secspikbos}}
Bosonic fluctuations in the ${\rm S}^5$ sector for any string solution in ${\rm AdS}_3 \x {\rm S}^1$ are 
massive fields with masses $\pm \mu$.  
From \eqref{spimu} we find the Lagrangian for the quadratic fluctuations in the ${\rm S}^5$ sector, $b_i~(i=1,\dots,4)$, is 
\be
	\mathcal{L} = 2 \sum_{i=1}^4 \left( \partial_+b_i \partial_- b_i - \frac{\alpha ^2 \nu ^2-D^2}{\alpha ^2-\beta ^2} b_i^2 \right)\,. \label{spifres5}
\ee
In this sector all fluctuations have the constant masses. 

The procedure for deriving the physical part of the fluctuated Lagrangian in the ${\rm AdS}_5$ sector is the same as that in the $(S,J)$ folded string case and the circular string case; 
introduce the fluctuation fields by \eqref{etap}, \eqref{etao} and \eqref{dgau},
integrate out the diagonal parts of the gauge field fluctuations, 
and then, use the $H$ gauge freedom such that physical fluctuations decouple from unphysical fluctuations.\footnote{
It should be noted that we use $v$ in \eqref{thespik} with \eqref{pmtheta} for $v$ in the expression \eqref{dgau}.
} 
Consequently we find the Lagrangian containing $a_1$ and $a_2$,
\be \label{spil1}
\mathcal{L}_1=2\sum _{i=1,2}\left( \partial_+a_i \partial_- a_i -\frac{ M_2+M_3 \alpha ^2 }{\alpha ^2-\beta ^2}a_i^2\right)\,.
\ee
As shown for the previous classical solutions, 
one shortcut way to the Lagrangian containing $a_3$ and $a_4$ is to perturb the tanh model Lagrangian directly. 
For the spiky string solution this approach has a big advantage because of the complicated expression for the classical solution. 
Substituting the classical solution \eqref{pmtheta}, \eqref{phispik} into the perturbed Lagrangian of the tanh model \eqref{quadtanh}, we obtain the Lagrangian for $a_3$ and $a_4$, 
\be \label{spil2} \ba{c}
\mathcal{L}_2= 2\bigg[ \partial _- a_3 \partial _+ a_3
+A_{33} a_3 ^2 +\partial _- a_4 \partial _+ a_4 +A_{44} a_4 ^2
+\left( A_{+}\partial _+ a_4+A_{-} \partial _- a_4\right) a_3 +A_{34}a_3a_4
\bigg]
 \,,
\ea\ee 
where 
\be\ba{rl}
&A_{33}= \frac{\alpha ^4\left(M_2-3 M_3 \alpha ^2\right) }{M_2^4 \left(\alpha ^2-\beta ^2\right)}   \bigg[ w_0^2 w_1^2 (C_0^4 +C_1^4) +2 C_0 C_1  w_0 w_1 (C_0^2+C_1^2) \left(w_0^2+w_1^2\right)  + C_0^2 C_1^2 \left(w_0^4+4 w_0^2 w_1^2+w_1^4\right) \\ 
& \bigg. \hspace{70pt }+(C_0^2 w_1^2 +C_1^2 w_0^2) \left(-M_1^2 \alpha^2 +2 w_0^2 \left( M_2+ M_3  \alpha ^2  \right) -w_0^4 \alpha ^2\right) +w_0^2 w_1^2 \left(D^2-M_3 \alpha ^2\right)^2 \\
& \bigg. \hspace{70pt }  + 2 C_0  C_1 w_0 w_1 \left(-M_1^2 \alpha^2 + (M_2+M_3 \alpha^2 ) \left(w_0^2+w_1^2\right) -w_0^2 w_1^2 \alpha ^2\right) 
 \  \bigg] -\frac{ M_2+M_3 \alpha ^2}{\alpha ^2-\beta ^2}  \,, 
\ea\ee
\be\ba{rl}
& A_{44} = \ -\frac{\left(\alpha ^2 \nu ^2-D^2\right)^2}{M_2^2 M_3^2 \left(\alpha ^2-\beta ^2\right)}  \bigg[ C_0^2 \left(M_1-w_0^2\right) \left(w_0^2-w_1^2\right)  - C_1^2 \left(M_1-w_1^2\right) \left( w_0^2-w_1^2\right) \\ 
 & \hspace{170pt } + (M_1^2 -M_1 \left(w_0^2+w_1^2\right) +w_0^2 w_1^2  ) (M_3 \alpha ^2-D^2 )    \bigg] \,, \hspace{15pt} 
\ea\ee
and
\be \ba{rl}
& A_{34}=\frac{4 \alpha ^2 \left(w_0^2-w_1^2\right) \left(\alpha ^2 \nu ^2-D^2\right)^{3/2} }{M_2^3 M_3 \left(\alpha ^2-\beta ^2\right)^{3/2}}\text{  }\sqrt{\left(M_3 \alpha ^2-D^2\right) r_1^4+\left(C_0^2-C_1^2+M_3 \alpha ^2-D^2\right) r_1^2-C_1^2} \\ 
& \hspace{25pt} \times \ \bigg[  w_0 w_1 \beta (C_0^2 +C_1^2 +M_3\alpha^2 - D^2) +C_0 C_1  \beta ( w_0^2+w_1^2)    
+\alpha^2 (M_1-w_0 w_1) (C_1 w_0+ C_0 w_1)  \bigg]  \,, 
\ea\ee
\be\label{a34pm}\ba{rl}
& A_{\pm }= \frac{ 2 \alpha (\alpha \mp \beta ) }{M_2 (\alpha \pm \beta )}\text{  }\sqrt{\frac{\alpha ^2 \nu ^2-D^2}{M_2^2 \left(\alpha ^2-\beta ^2\right)}} \, \bigg[  w_0 w_1 \alpha (C_0^2+C_1^2)  +w_0 w_1 \alpha  \left(M_3 -D^2 \alpha ^2\right) \\
& \hspace{140pt} \mp (C_1 w_0+ C_0 w_1) \left(M_1 \alpha^2-w_1^2 \alpha ^2\right)  +  C_0 C_1 \alpha \left(w_0^2+w_1^2\right)  \bigg] \,.
\ea
\ee 

Finally we shall consider the case of no stretching in ${\rm S}^5$, which is achieved by taking the limit $D,\nu \to 0$, and correspondingly, $M_2\to \alpha^2 M_1$, $M_3\to M_1$. 
In this case the fluctuations in the ${\rm S}^5$ \eqref{spifres5} become massless. 
In the ${\rm AdS}_5$ sector we have the following Lagrangian, 
\be\label{padsbosspi}
\mathcal{L}=2 \left[ \sum _{i=1,2}\left( \partial_+a_i \partial_- a_i -\frac{2\alpha^2 M_1}{\alpha ^2-\beta ^2}a_i^2\right) + \partial_+a_3 \partial_- a_3 -\frac{2f^2 \left( w_0^2-w_1^2\right)^2-2 \alpha ^2 \left(M_1^2+w_0^2 w_1^2\right) }{M_1 \left(\alpha ^2-\beta ^2\right)} a_3^2+\partial_+a_4 \partial_- a_4 \right] \,,  
\ee
where we have rewritten $C_0$ and $C_1$ as $C_0=w_1 f$ and $C_1=-w_0 f$, respectively, such that they solve the second Virasoro constraint \eqref{spvir2} with $D=\nu = 0$. 
Hence it turns out that the $D,\nu\to 0$ limit is well defined at the level of the fluctuation Lagrangian in the bosonic sector . 

Because the folded string solution in pure ${\rm AdS}_3$ is realized as a special case of the spiky string string solution, 
the Lagrangian \eqref{padsbosspi} should recover the fluctuation Lagrangians for the folded string, \eqref{a1a2nu0} and \eqref{a3a4}. 
In fact the folded string solution corresponds to the limit $M_1 \to \rho^2$, $w_0\to \kappa$, $w_1 \to w$ and $\alpha,\beta \to 0$, 
and one can see that the Lagrangian \eqref{padsbosspi} reduces to the sum of \eqref{a1a2nu0} and \eqref{a3a4} in this limit.

\subsection{Fermionic fluctuations in reduced theory\label{secspikfer}}

Once given the component fields of fermionic fluctuations by \eqref{ferRL1} - \eqref{ferRL5}, we can write down the fermionic fluctuation Lagrangian in terms of $\alpha_i$ and $\beta_i$,\footnote{
Again we should use $v$ in \eqref{thespik} with \eqref{pmtheta} for $\delta \Psi_{_L}$ instead of the original $v$ for the $(S,j)$ folded string.
}   
\be \label{ferspik}\ba{rl}
	{\cal L}_{\rm F} \ = \! &2 \Big[ \sum _{i=1}^8\left( \alpha_i \dm \alpha_i +\beta_i \dm \beta_i \right) + A_{\alpha\alpha} \left(-\alpha_1 \alpha_2 - \alpha_3 \alpha_4 -\alpha_5 \alpha_6 +\alpha_7 \alpha_8  \right) \\ 
	&
	 + A_{\beta\beta} \left(- \beta_1 \beta_2-\beta_3 \beta_4-\beta_5 \beta_6+\beta_7 \beta_8 \right) \\ 
	& 
	+ A_{\alpha\beta} \left( -\alpha_3\beta_1  -\alpha_1\beta_3  +\alpha_7\beta_5  -\alpha_5\beta_7  +\alpha_4\beta_2  +\alpha_2\beta_4 + \alpha_8\beta_6 - \alpha_6\beta_8  \right)  \Big] \,.
\ea\ee
where 
\be\ba{rl}
& A_{\alpha\alpha}=-\frac{1}{2}A_+ \,, ~~~~~
 A_{\beta\beta}=\frac{1}{2}A_- \,,\Big.  \\
& A_{\alpha\beta} = 2 \frac{M_2+\sqrt{M_2 M_3} \alpha }{\sqrt{\left(M_2 + 2 \sqrt{M_2 M_3} \alpha +M_3 \alpha ^2\right) \left(\alpha ^2-\beta ^2\right)}} \,.
\ea\ee
$A_\pm$ are defined in the bosonic sector, \eqref{a34pm}.

The Lagrangian \eqref{ferspik} again describes four decoupled systems, each of which has four fermionic component fields.
After rotating the fermionic component fields, as discussed in the folded string case in section \ref{foldfer}, 
we have four copies of the following Lagrangian, 
\be \la{spikferla}
\mathcal{L}_{\rm f}= \bar\Psi \gamma^a \partial_a\Psi  - \frac{M_2+\sqrt{M_2 M_3} \alpha }{\sqrt{\left(M_2+2 \sqrt{M_2 M_3} \alpha +M_3 \alpha ^2\right) \left(\alpha ^2-\beta ^2\right)}}\bar\Psi\Gamma_2\Psi \,,
\ee
where $\Psi$ is a real four component spinor. 
 
Let us now discuss the special case where the string is in pure ${\rm AdS}_3$, 
that is, $D, \mu \to 0$, $M_2\to \alpha^2 M_1$ and $M_3\to M_1$. 
Recalling that we are considering that case of $M_1>0$, 
then we find the coefficient of the mass term of the fermionic fluctuations becomes 
\be
 \sqrt{\frac{\alpha ^2M_1}{\alpha^2-\beta^2}} \,.
\ee
Hence the $D, \mu \to 0$ limit is well-defined at the level of the fluctuation Lagrangian in the fermionic sector.

For consistency the fermionic Lagrangian \eqref{spikferla} should recover the fermionic fluctuation Lagrangian for the folded string in pure ${\rm AdS}_3$. 
Taking the corresponding limit $M_1 \to \rho^2$, $w_0\to \kappa$, $w_1 \to w$ and $\alpha,\beta \to 0$ yields the mass term with a coefficient $ \rho'$ which agrees with \cite{ft1}. 
Hence the fermionic part of the quadratic fluctuations around the folded string without the ${\rm S}^5$ sector is recovered.

\renewcommand{\theequation}{6.\arabic{equation}}
 \setcounter{equation}{0}

\section{Folded string with orbital momentum and winding in ${\rm S}^1$ of ${\rm S}^5$\label{mwfold}}

In this section we shall study semiclassical expansion around the generalized folded string solution with both the orbital momentum and the
winding in ${\rm S}^1$ of ${\rm S}^5$ which was constructed in \cite{grrtv}.  
After showing how to achieve the generalized homogeneous folded string from the $(S,J)$ spiky string, we will evaluate the characteristic frequencies for the quadratic fluctuations for bosons in \ref{secmwbos} and for fermions in \ref{secmwfer} in order to compare them with the result of \cite{grrtv}. 

The generalized folded string solution also has the open string counterpart, a null cusp solution, which is obtained by a combination of analytic continuation on the worldsheet and $SO(2,4)$ rotation from the generalized folded string solution \cite{grrtv}.  
Because the $SO(2,4)$ symmetry is obscure in the reduced theory, the generalized folded string and its open string counterpart are connected by the analytic continuation in the reduced theory. 
Hence the equivalence between the two classical solutions becomes rather trivial by the Pohlmeyer reduction.

Because of the homogeneous nature of the generalized folded string, its fluctuation Lagrangian has constant coefficients, and thus, quantum corrections to the 
partition function can be computed. 
In fact, in \cite{grrtv}, the expansion around the null cusp solutions are carried out and, the one-loop and two-loop corrections are determined. 
Directly from the equivalence of the generalized folded string and the null cusp solution, these quantum corrections are also relevant to the folded string. 

Below we shall evaluate the characteristic frequencies of the quadratic fluctuations and compare them with the one loop computation in \cite{grrtv}.
Reflecting the fact the generalized folded string has both the orbital momentum and the winding in the ${\rm S}^1$ sector, 
this is another case of $\mu_+\ne \mu_-$ in the reduced theory. 
Therefore the conclusion of the one loop computation is very similar to the $(S,J)$ circular string case in section \ref{sjcs}; two of eight bosonic frequencies which can be derived by the perturbation in the complex sinh-Gordon model do not agree with the string theory result. But the the discrepancy is removed by considering the fermionic contributions, and the total sum of the frequencies agrees with the string theory result. 
In order to show this we shall find a 2d Lorentz transformation such that all of the bosonic and fermionic frequencies become the same as the frequencies of the corresponding fluctuations in the original theory in \ref{secmn2d}. 

We shall first review the generalized folded string solution with both the orbital momentum and the
winding in ${\rm S}^1$ of ${\rm S}^5$. 
Introduce the global coordinates in ${\rm AdS}_3 \x {\rm S}^1$ by  
\be
 ds^2=-\cosh ^2\rho dt^2+d\rho ^2+\sinh ^2\rho d\theta ^2+d\vp^2\,,
\ee
then we expressed the solution in the conformal gauge, 
\be \la{gmsoln} 
    t=\kappa\tau \,,~~~~~ \rho= \rho(\sigma) \, , \ \ \ \ \ \theta =\kappa\tau  + \vartheta(\sigma) \, , \ \ \ \ \ \ 
   \vp=\nu \tau  + m \s  \,,
\ee
where
\be \label{gmcond} \ba{l}
   \cosh \rho(\s) =
    \sqrt{1 + \gamma^2 }      \cosh (  \ell\sigma)  \, , ~~~~ 
 \tan \vartheta (\s)  =  \gamma  \coth (  \ell\sigma) \, , ~~~~
 \gamma \equiv {\nu m \ov \k \ell} \, ,  \\
\kappa^2=\ell^2+\nu^2 + m^2  \, .   
\ea\ee
Here we are considering the following limit, 
\be
\ell \gg 1    \, , ~~~~  \nu  \gg 1 \  \, ,   ~~~~
 \k,\ell,\nu, m   \gg 1  \, ,  ~~~~
 { \nu \ov \ell} = { \rm fixed} \, ,~~~~      
 { m  \ov \ell} = { \rm fixed}   \, .
\ee
In this solution only three parameters are independent. We will use $\kappa$, $\nu$ and $m$ when we calculate quadratic fluctuations.
If we set the winding in a large circle of ${\rm S}^5$, $m$, to be zero, the solution reduces to the $(S,J)$ folded string solution discussed in section \ref{secfold}.

To compare this solution with the $(S,J)$ spiky string solution in section \ref{sjspik}, we rewrite the generalized folded string solution in terms of another radial coordinate $v$ introduced in \eqref{rtovtorho} and derive the equation for $v'$,  
\be
v'=-2 \ell v \sqrt{(1+v) \left(1-v-2 \gamma^2 v\right)} \,,
\ee
which shows
$v'$ vanishes at $v=-1,~0,~\frac{1}{1+2 \gamma^2}$ corresponding to the three roots $v_1$, $v_2$ and $v_3$ with $ v_1\leq0\leq v_2\leq v_3\leq 1$. 
Then we find that the generalized folded string solution is realized by taking the following limit in the $(S,J)$ spiky string solution,  
\be \label{sstogfs}
v_1=-1\,,~~~v_2=0\,,~~~ v_3=\frac{1}{1+2 \gamma^2}\,.
\ee
which is different form the three limiting cases considered in \cite{iktt}.\footnote{
The connection of the $n$-spike string and its long string limit was studied in \cite{krutir} in the context of recovering ${\rm AdS}_3 \x {\rm S}^1$ string solutions from the asymptotic $SL(2)$ Bethe Ansatz equations. For $n=2$ the limiting solution reduces to the generalized folded string solution.  
}

Let us next construct the corresponding solution in the reduced theory. 
For this purpose we express the generalized folded string solution \eqref{gmsoln} in terms of the embedding coordinates, 
\begin{equation}\label{gkpex}\ba{c}
Y_0+iY_{3}=\cosh \! \rho (\sigma ) \,e^{i\kappa \tau}\,, 
~~~ Y_1+iY_2= \sinh \! \rho (\sigma ) \,e^{i\vartheta (\sigma)}\,, ~~~
X_5+iX_6=e^{i(\nu \tau+m \sigma)}\,, 
\ea
\end{equation}
where $\rho (\sigma)$ and $\vartheta( \sigma)$ are given in \eqref{gmcond}, and related to the parameters in the ${\rm S}^1$ sector. 
The mass scale of the reduced theory $\mu$ can be extracted from $T_{++}^{\rm AdS}$ and $T_{--}^{\rm AdS}$. In the present case we have 
\be
	T_{\pm \pm }^{\rm AdS}=-(\nu \pm m )^2\,, 
\ee
which imply $\mu_\pm=\nu\pm m$, and then, $\mu$ should be introduced by their product, 
\be \label{mnmupm}
	\mu = \sqrt{\mu_+ \mu_-}= \sqrt{\nu ^2-m^2} \,.
\ee
As done in the semiclassical computation for the other solutions we use the tanh model in the reduced theory. 
Once we obtain the mass scale $\mu$, 
the solution in the original theory, \eqref{gkpex}, is encoded into a solution of the tanh model in the reduced theory by using \eqref{red}, 
\be \ba{l}
	\phi _{\!_A}=\frac{1}{2} \text{log}\left[\frac{2 \kappa ^2-\nu ^2-m^2+2 \sqrt{\left(\kappa ^2-m^2\right) \left(\kappa ^2-\nu ^2\right)}}{\nu ^2-m^2}\right] \,,\\
	\theta_{\!_A} =\frac{ \kappa ^2-m^2 }{\sqrt{\nu ^2-m^2}}\tau  \,,
\ea
\ee
Plugging these into \eqref{g0ta} and \eqref{gatan} we obtain the classical solution for the deformed gWZW model, 
\be
g_0=
\left(
\begin{array}{cccc}
 0 & vV_1 & -vV_2 & 0 \\
 - v^* V_1 & 0 & 0 &  v^* V_2 \\
 vV_2 & 0 & 0 & - v V_1 \\
 0 & -v^* V_2 &  v^* V_1 & 0
\end{array}
\right) \,, \\
\ee
where
\be \label{vvvmn}
\ba{l}
v=e^{\frac{i \left(\kappa ^2-m^2\right)}{\sqrt{\nu ^2-m^2}}\tau }\,, \\
V_1=\frac{ \kappa ^2-m^2+\sqrt{\left(\kappa ^2-m^2\right) \left(\kappa ^2-\nu ^2\right)} }{\nu ^2-m^2}\sqrt{\frac{\nu ^2-m^2}{2 \kappa ^2-\nu ^2-m^2+2 \sqrt{\left(\kappa ^2-m^2\right) \left(\kappa ^2-\nu ^2\right)}}} \,, \\
 V_2=-\frac{ \kappa ^2-\nu ^2+\sqrt{\left(\kappa ^2-m^2\right) \left(\kappa ^2-\nu ^2\right)} }{\nu ^2-m^2}\sqrt{\frac{\nu ^2-m^2}{2 \kappa ^2-\nu ^2-m^2+2 \sqrt{\left(\kappa ^2-m^2\right) \left(\kappa ^2-\nu ^2\right)}}} \,. 
\ea
\ee
and the corresponding classical gauge fields are $A_{\pm 0}=\frac{i}{2}\, a_{\pm 0}R_2$ with 
\be
\ba{c}
a_{+0}=\frac{m^2-2 \kappa ^2+\nu ^2}{ \sqrt{\nu ^2-m^2}} \,, \\
a_{-0}=- \sqrt{\nu ^2-m^2} \,.
\ea
\ee
With this choice, $g_0^{-1}\partial_+ g_0$ and $g_0^{-1}A_+ g_0$ are constants, and then, the physical part of the quadratic fluctuation Lagrangian has constant coefficients after properly choosing the $H$ gauge. 
Hence we can straightforwardly evaluate characteristic frequencies of the quadratic fluctuations, which should be compared with the frequencies in the original string theory.

\subsection{Bosonic fluctuations in reduced theory\label{secmwbos}}
Four physical modes in the ${\rm S}^5$ sector yield four bosonic fluctuations with the masses $\pm \mu$. 
In the present case we have $\mu^2 =  \nu ^2-m^2 $, then their frequencies are 
\be \label{gefos5}
	\pm \sqrt{n^2 +\nu ^2-m^2} \,,
\ee
which is consistent with \cite{grrtv}. 

For the ${\rm AdS}_5$ sector one easy way to obtain the constant coefficient Lagrangian for physical fluctuations is again to introduce the component fields of $\eta$ and $\delta A_\pm$ as in \eqref{etap}, \eqref{etao} and \eqref{dgau}, and then, to use the $H$ gauge symmetry such that the physical fields decouple from the unphysical fields as in the folded string case.\footnote{
The original $v$ in \eqref{dgau} should be replaced by $v$ in \eqref{vvvmn}.
} 
Two of the bosonic fluctuations in the ${\rm AdS}_5$ sector are described by the Lagrangian, 
 \be 
\mathcal{L}_1=2\sum _{i=1,2}\left( \partial_+a_i \partial_- a_i -(2 \kappa ^2-\nu ^2-m^2)a_i^2\right)\,,
\ee
Then their frequencies are 
\be\label{gefoa1a2}
 \pm \sqrt{n^2+2 \kappa ^2-\nu ^2-m^2} \,,
\ee
which agree with the result in \cite{grrtv}. The problem appears in the sector of the other two bosonic fluctuations, which can also be captured directly by the perturbation in the complex sinh-Gordon model, 
\be  \ba{c}
\mathcal{L}_2= 2\bigg[ \partial _- a_3 \partial _+ a_3
-4 \left(\kappa ^2-\nu ^2\right) a_3 ^2 +\partial _- a_4 \partial _+ a_4 
-2 \sqrt{\nu^2 -m^2} \left( \partial _+ a_3+ \partial _- a_3\right) a_4
\bigg]
 \,.
\ea\ee
The equation for the characteristic frequencies is 
\be \label{genbored}
	\Omega^4 - 2 \left( n^2+2 \kappa ^2-2 m^2  \right) \Omega^2+n^2 \left(n^2+4 \kappa ^2-4 \nu ^2\right) =0\,,
\ee
which is not the same as the corresponding equation in \cite{grrtv}, 
\be \label{genboorg}
	\Omega^4-2 \Omega^2 \left(n^2+2 \kappa ^2\right)+8 \nu m n \Omega  +n^2 \left( n^2+4 \kappa ^2-4 \nu ^2-4 m^2 \right) =0\,.
\ee
Therefore characteristic frequencies of the reduced theory and the original theory are different in this sector.  
However, this does not imply that our conjecture on the quantum equivalence between the original string theory and the reduced theory, \eqref{eqpa}, breaks down. 
Later we will show this discrepancy should cancel with that of fermionic fluctuations by finding a specific Lorentz transformation making all of the frequencies 
the same as those in the original string theory.

\subsection{Fermionic fluctuations in reduced theory\label{secmwfer}}
The parameterization for the fermionic fields \eqref{ferRL1} - \eqref{ferRL5} in the fermionic part of the fluctuation Lagrangian \eqref{quafluc} gives the constant coefficient Lagrangian,\footnote{
The original $v$ in $\delta \Psi_{_L}$ should be replaced by $v$ in \eqref{vvvmn}.
}  
\be\ba{l}
	{\cal L}_{f}=2 \Big[ \sum _{i=1}^8\left( \alpha_i \dm \alpha_i +\beta_i \dm \beta_i \right) \nonumber \\ \hspace{15pt}  +\sqrt{\nu ^2-m^2} \left(\alpha_1 \alpha_2 + \alpha_3 \alpha_4 +\alpha_5 \alpha_6 -\alpha_7 \alpha_8  - \beta_1 \beta_2-\beta_3 \beta_4-\beta_5 \beta_6+\beta_7 \beta_8 \right)\nonumber \\ \hspace{15pt}
	+\frac{2 \left(-m^2+\kappa ^2+\sqrt{\left(-m^2+\kappa ^2\right) \left(\kappa ^2-\nu ^2\right)}\right)}{\sqrt{-m^2+2 \kappa ^2-\nu ^2+2 \sqrt{\left(-m^2+\kappa ^2\right) \left(\kappa ^2-\nu ^2\right)}}}  \left( \alpha_1 \beta_4 + \alpha_2 \beta_3 - \alpha_3 \beta_2 -\alpha_4 \beta_1 -\alpha_5 \beta_8 +\alpha_6 \beta_7  +\alpha_7 \beta_6  
	-\alpha_8 \beta_5 \right)  \Big] \,.
\ea\ee
The fermionic characteristic frequencies are given by solving the following equation, 
\be \label{genfered}
\Omega^4- \frac{1}{2} \left(4 n^2+4 \kappa ^2+\nu ^2-5 m^2\right) \Omega^2 +\frac{1}{16} \left( 4 n^2+4 \kappa ^2-\nu ^2 -3 m^2\right)^2 =0 \,,
\ee
which is not the same as the equation for the fermionic frequencies in the original theory, 
\be \label{genfeorg} \ba{c}
\Omega^4  - \frac{1}{2}  \left( 4 n^2+4 \kappa ^2+\nu ^2 -3 m^2 \right) \Omega^2 + 2 n \nu m \Omega \hspace{100pt} \\ \hspace{20pt}
+\frac{1}{16} \left[ 9 m^4+\left(4 n^2+4 \kappa ^2-\nu ^2\right)^2 - 2 m^2 \left( 20 n^2 +12 \kappa ^2-3\nu ^2 \right)\right] =0 \,,
\ea\ee
and consequently, the characteristic frequencies do not match.

\subsection{2d Lorentz boost\label{secmn2d}}
In section \ref{cirsumf} we carried out the large $\cal J$ expansion and showed the sum of the frequencies in the reduced theory is the same as that in the original string theory up to the order $1/{\cal J}^2$. 
Because the sum of the frequencies should be invariant under 2d Lorentz boost on the worldsheet, the sums of the frequencies of these two theories are the same if all of the frequencies in the reduced theory become the same as the corresponding frequencies in the original theory by a single 2d Lorentz boost.   
The $(S,J)$ circular string is the case. 

The present situation is very similar. Six of bosonic frequencies, \eqref{gefos5} and \eqref{gefoa1a2}, are Lorentz invariant, while the others frequencies  described by \eqref{genbored} and \eqref{genfered} are changed by the Lorentz transformation.  
Here we look for a 2d Lorentz transformation which applies for the equations for both the bosonic and fermionic frequencies.
If we introduce the 2d Lorentz transformation in the reduced theory by 
\be \label{mn2dboost}
	\tau \to p_2 \tau + q_2 \sigma \,, ~~~\sigma \to q_2 \tau + p_2 \sigma \,, ~~~p_2^2-q_2^2=1\,,
\ee
with
\be \label{mn2dboostt}
	p_2=\frac{\nu }{\sqrt{\nu ^2-m^2}}\,,~~~~~~q_2=-\frac{m}{\sqrt{\nu ^2-m^2}}\,,
\ee
then \eqref{genbored} and \eqref{genfered} become \eqref{genboorg} and \eqref{genfeorg}, respectively.
Hence the frequencies match for all the fluctuations.
This implies the total sum of the quantum corrections to the partition function in the reduced theory agree with the string theory result, and so, supports our conjecture in \eqref{eqpa}.

It is worth mentioning that the transformation \eqref{mn2dboost}, \eqref{mn2dboostt} is exactly the same as the one setting $\mu_+=\mu_-$ in \eqref{mnmupm}. 
In fact, applying the $2d$ Lorentz boost to the generalized folded string solution \eqref{gkpex}, we finds that the stress tensor becomes  
\be
 T_{\pm \pm }^{\rm AdS}=-(\nu^2  -m^2 ) \,,
\ee
which implies $\mu_+=\mu_-=\nu^2  -m^2$. 
This is the same observation as in the circular string case in section \ref{sjcs}.

\section{Concluding remarks}
In this paper we have discussed the semiclassical expansions around several classical string configurations localized in ${\rm AdS}_3 \times {\rm S}^1$ using the Pohlmeyer-reduced form of the ${\rm AdS}_5 \times {\rm S}^5$ superstring theory. 

To avoid the difficulty in reducing a classical string solution, i.e., fixing the $G$ gauge and the $H \times H$ gauge, we have used the embedding of the complex sinh-Gordon model into the deformed WZW model, which allows us to easily understand inhomogeneous string solutions through the reduced theory, in particular, to uncover the relation between the fluctuations in the Pohlmeyer-reduced form and those derived by perturbing the Nambu action in the original string theory.
Another result is that if $\mu_+\ne \mu_-$, that is, if string has both the winding and the orbital momentum in ${\rm S}^1$ of ${\rm S}^5$, 
the perturbation in the reduced theory gives the equivalent set of characteristic frequencies in the sense that the sum of the frequencies agrees with the string theory result, although the discrepancy is found for the frequencies of the individual fluctuations. 
The agreement of the individual fluctuation frequencies is achieved by applying the $2d$ transformation such that it sets $\mu_+=\mu_-$ in the classical solutions.  
We have also derived the fluctuation Lagrangian for the $(S,J)$ spiky string solution which is expected to agree with the original superstring theory.  

One can apply the technique used in the present paper to string solutions in ${\rm R}\x {\rm S}^3$, where a homogeneous case was studied in the reduced theory in \cite{hit}. 
However, if we try to understand strings in larger spacetime, such as ${\rm AdS}_5\x {\rm S}^1$ and ${\rm R}\x {\rm S}^5$, we would need to follow different strategy in reducing a string solution to a reduced-theory solution and integrating out unphysical fields.  

In this paper we have shown that the limit of no stretching in the ${\rm S}^1$, corresponding to the $\mu \to 0$ limit, is well-defined at the level of the fluctuation Lagrangian of the reduced theory. 
So it would be interesting to check if we can extend the one-loop equivalence \eqref{eqpa} to the two-loop level and take the $\mu \to 0$ limit at the level of the effective Lagrangian. 
It is still an open problem how this limit should be taken at the level of the classical Lagrangian \cite{gt2}.

\section*{Acknowledgements}

We would like to thank A. Tseytlin for suggesting the problems and many discussions. We also thank B. Hoare and R. Roiban for useful discussions and comments. 
YI is supported by the Ishizaka Foundation.

\appendix

\section{$\mathfrak{psu}(2,2|4)$ superalgebra\label{appsupcos}}

In this appendix we summarize the $\mathfrak{psu}(2,2|4)$ superalgebra and its parameterization.\footnote{
In this paper we follow the notation of \cite{gt1}. 
} 
We shall start with the $\mathfrak{su}(2,2|4)$ superalgebra as it is spanned by $8 \times 8$ supermatrices $\mathfrak{f}$. 
In general $\mathfrak{f}$ is expressed in terms of $4\times 4$ matrices, 
\begin{eqnarray}
	\mathfrak{f}= \left(
	\begin{array}{@{\,}cc@{\,}}
		\mathfrak{A} & \mathfrak{X} \\
		\mathfrak{Y} & \mathfrak{D}
	\end{array}	  
	\right) \,,
\end{eqnarray}
where the matrices $\mathfrak{A}$, $\mathfrak{D}$ are Grassmann even and $\mathfrak{X}$, $\mathfrak{Y}$ are Grassmann odd. 
The superalgebra $\mathfrak{su}(2,2|4)$ requires that 
the matrix $\mathfrak{f}$ should vanish by taking the supertrace, 
\begin{eqnarray}
	{\rm Str} \mathfrak{f}\equiv {\rm tr}\mathfrak{A}-{\rm tr}\mathfrak{D}=0\,,  \label{strc}
\end{eqnarray}
and satisfy the reality condition, 
\begin{eqnarray}
	\mathfrak{f}^{\dagger}H+H\mathfrak{f}=0\,, \label{realc}
\end{eqnarray}
with a Hermitian matrix $H$. 
It is convenient to choose $H$ to be the following form,  
\begin{eqnarray} \label{hdef}
	H= \left(
	\begin{array}{@{\,}cc@{\,}}
		\Sigma & 0 \\
		0 & {\bf 1}
	\end{array}	  
	\right) \,,
\end{eqnarray}
where $\Sigma$ is expressed as 
\begin{eqnarray}
	\Sigma= \left(
	\begin{array}{@{\,}cccc@{\,}}
		1 & 0 & 0 & 0 \\
		0 & 1 & 0 & 0 \\
		0 & 0 & -1 & 0 \\
		0 & 0 & 0 & -1 
	\end{array}	  
	\right) \,, \label{sigmdea}
\end{eqnarray}
and ${\bf 1}$ is the $4 \x 4$ identity matrix. 
With this choice of $H$, the reality condition \eqref{realc} gives the relations,
\begin{eqnarray}
	\mathfrak{A}^{\dagger}=-\Sigma \mathfrak{A} \Sigma \,,~~~
	\mathfrak{D}^{\dagger}= -\mathfrak{D}\,,~~~
	\mathfrak{Y}=-\mathfrak{X}^{\dagger}\Sigma \,.
\end{eqnarray}
So we see that the bosonic matrix $\mathfrak{A}$ belongs to $\mathfrak{u}(2,2)$ and the other bosonic matrices $\mathfrak{D}$ belong to $\mathfrak{u}(4)$. 
The only one combination of each $\mathfrak{u}(1)$ generator, $i{\bf 1}$, satisfies the reality condition \eqref{realc} and 
supertraceless condition \eqref{strc}. 
Hence the bosonic subalgebra of $\mathfrak{su}(2,2|4)$ is decomposed as 
\begin{eqnarray}
	\mathfrak{su}(2,2) \oplus \mathfrak{su}(4) \oplus \mathfrak{u}(1)\,. 
\end{eqnarray}
The superalgebra $\mathfrak{psu}(2,2|4)$ is defined as the quotient algebra of $\mathfrak{su}(2,2|4)$ over this $\mathfrak{u}(1)$ factor.

One important property of the $\mathfrak{psu}(2,2|4)$ superalgebra is that it admits a $\mathbb{Z}_4$ automorphism such that the condition 
$\mathbb{Z}_4(\mathfrak{f})=\mathfrak{f}$ determines the subgroup $G=Sp(2,2)\times Sp(4)$ of $F=PSU(2,2|4)$. 
Define the automorphism $\mathfrak{f} \to \Omega (\mathfrak{f})$ by 
\begin{eqnarray}
	\Omega (\mathfrak{f}) = -\left(
	\begin{array}{@{\,}cc@{\,}}
		K\mathfrak{A}^{\rm t}K & -K\mathfrak{Y}^{\rm t}K \\
		K\mathfrak{X}^{\rm t}K & K\mathfrak{D}^{\rm t}K
	\end{array}	  
	\right) \,,
\end{eqnarray}
where the $4\times 4$ matrix $K$ is chosen to be 
\begin{eqnarray}\label{kdef}
	K = \left(
	\begin{array}{@{\,}cccc@{\,}}
		0 & -1 & 0 & 0 \\
		1 & 0 & 0 & 0 \\
		0 & 0 & 0 & -1 \\
		0 & 0 & 1 & 0 
	\end{array}	  
	\right) \,,
\end{eqnarray}
which satisfies $K^2 =-{\bf 1}$, 
then any matrix in $\mathfrak{psu}(2,2|4)$ can be decomposed as 
\begin{eqnarray}
	\mathfrak{f}=\mathfrak{f}_0 \oplus \mathfrak{f}_{1}\oplus \mathfrak{f}_2 \oplus \mathfrak{f}_3\,, 
\end{eqnarray}
where $\mathfrak{f}_k$ are eigenstates of $\Omega$, 
\begin{eqnarray}
	 \Omega (\mathfrak{f}_k)=i^k\mathfrak{f}_k \,, 
\end{eqnarray}
and given by 
\begin{eqnarray}
	&& \mathfrak{f}_0=\frac{1}{4}\left( \mathfrak{f}+ \Omega (\mathfrak{f}) + \Omega ^2(\mathfrak{f})+ \Omega ^3(\mathfrak{f}) \right) 
	= \frac{1}{2}\left(
	\begin{array}{@{\,}cc@{\,}}
		\mathfrak{A}-K\mathfrak{A}^{\rm t}K & 0 \\
		0 & \mathfrak{D}-K\mathfrak{D}^{\rm t}K
	\end{array}	  
	\right) \,, \nonumber \\
	&& \mathfrak{f}_1=\frac{1}{4}\left( \mathfrak{f}-i \Omega (\mathfrak{f}) - \Omega ^2(\mathfrak{f})+ i\Omega ^3(\mathfrak{f}) \right) 
	= \frac{1}{2}\left(
	\begin{array}{@{\,}cc@{\,}}
		0 & \mathfrak{X}-iK\mathfrak{Y}^{\rm t}K \\
		\mathfrak{Y}+iK\mathfrak{X}^{\rm t}K & 0
	\end{array}	  
	\right) \,, \nonumber \\
	&& \mathfrak{f}_2=\frac{1}{4}\left( \mathfrak{f}- \Omega (\mathfrak{f}) + \Omega ^2(\mathfrak{f})- \Omega ^3(\mathfrak{f}) \right) 
	= \frac{1}{2}\left(
	\begin{array}{@{\,}cc@{\,}}
		\mathfrak{A}+K\mathfrak{A}^{\rm t}K & 0 \\
		0 & \mathfrak{D}+K\mathfrak{D}^{\rm t}K
	\end{array}	  
	\right) \,, \nonumber \\
	&& \mathfrak{f}_3=\frac{1}{4}\left( \mathfrak{f}+ i\Omega (\mathfrak{f}) - \Omega ^2(\mathfrak{f})-i \Omega ^3(\mathfrak{f}) \right) 
	= \frac{1}{2}\left(
	\begin{array}{@{\,}cc@{\,}}
		0 & \mathfrak{X}+iK\mathfrak{Y}^{\rm t}K \\
		\mathfrak{Y}-iK\mathfrak{X}^{\rm t}K & 0
	\end{array}	  
	\right) \,. \nonumber \\
	~
\end{eqnarray}
They satisfy the following commutation relation, 
\be  \label{fij}
	\left[ \mathfrak{f}_i,\mathfrak{f}_j \right] \subset \mathfrak{f}_{i+j~{\rm mod}~4} \,.
\ee
Let us denote $\mathfrak{g}=\mathfrak{f}_0$ and $\mathfrak{p}=\mathfrak{f}_2$. 
Then $\mathfrak{g}$ is the algebra of the subgroup $G$ of $F$ defining the coset $F/G$, and $\mathfrak{p}$ corresponds to the bosonic coset component. 
The other two components, $ \mathfrak{f}_1$, $ \mathfrak{f}_3$, are fermionic parts.

Now let us discuss a further $\mbb{Z}_2$ decomposition, 
which defines the group $H$ and the coset $G/H$.  
Here we shall introduce an element $T$ of the maximal Abelian subalgebra of $\mathfrak{p}$ by 
\be \label{tdef}
T=\fr{i}{2}\trm{ diag}\left(1,\, 1,\, -1,\, -1,\, 1,\, 1,\, -1,\, -1\right)\,.\ee
The $\mbb{Z}_2$ decomposition is then given by
\be\mf{f}^\parallel_r=-\left[T,\left[T,\mf{f}_r\right]\right]\,,\hs{40pt} \mf{f}^\perp_r=-\{T,\{T,\mf{f}_r\}\}\,.\ee
It should be noted that this is an orthogonal decomposition, that is
\be\ba{c}
\mf{f}=\mf{f}^\parallel\oplus\mf{f}^\perp\,,\\
\trm{STr}(\mf{f}^\parallel \mf{f}^\perp)=0\,.\ea
\ee
and they form the following commutation relation, 
\be \label{fpepa}
\left[\mf{f}^\perp,\,\mf{f}^\perp\right]\subset\mf{f}^\perp\,,\hs{30pt}
\left[\mf{f}^\perp,\,\mf{f}^\parallel\right]\subset\mf{f}^\parallel\,,\hs{30pt}
\left[\mf{f}^\parallel,\,\mf{f}^\parallel\right]\subset\mf{f}^\perp\,.\ee
Identify $\mf{h}=\mf{f}_0^\perp$, $\mf{m}=\mf{f}_0^\parallel$, $\mf{a}=\mf{f}_2^\perp$, 
$\mf{n}=\mf{f}_2^\parallel$. 
In fact $\mf{a}$ is the maximal Abelian subspace of $\mf{p}$, and the algebra $\mf{h}$ of the subgroup $H$ of $G$ is defined as the stabilizer of $T$ in $\mf{g}$, i.e., $[ \mf{h},T]=0$. 
Together with the commutation relations \eqref{fij} and \eqref{fpepa}, one finds 
these elements satisfy 
\be
\left[\mf{a},\mf{a}\right]\subset 0\,, \hs{20pt}
\left[\mf{a},\mf{h}\right]\subset 0\,, \hs{20pt}
\left[\mf{h},\mf{h}\right]\subset \mf{h}\,, \hs{20pt}
\left[\mf{m},\mf{m}\right]\subset \mf{h}\,, \hs{20pt}
\left[\mf{m},\mf{h}\right]\subset \mf{m}\,, \hs{20pt}
\left[\mf{m},\mf{a}\right]\subset \mf{n}\,, \hs{20pt}
\left[\mf{n},\mf{a}\right]\subset \mf{m} \,. \ee

For the specific choice of the matrices $H$, $K$ and $T$ in \eqref{hdef}, \eqref{kdef} and \eqref{tdef}, respectively, then one can uniquely express  
general elements of $\mf{m}$, $\mf{h}$, $\mf{a}$, $\mf{n}$, $\mf{f}_1^{\parallel}$, $\mf{f}_1^{\perp}$, $\mf{f}_3^{\parallel}$ and $\mf{f}_3^{\perp}$ 
in terms of their matrix components.  
The following four components are relevant when we determine the fluctuation Lagrangian of the reduced theory. 
The subspace $\mf{m}$ of $\mf{g}$ corresponds to physical fluctuations in the reduced theory, 
\be
\ba{c}
\mf{m}=\left( \ba{cc}
	\mf{m}_{_A} & 0 \\
	0 & \mf{m}_{_S}
\ea \right) \,, \\
\mf{m}_{_A} =\left(
	\begin{array}{cccc}
	0 & 0 & a_{1}+ia_{2}& a_{3}+ia_{4}\\
	0 & 0 & a_{3}-ia_{4}& -a_{1}+ia_{2}\\
	a_{1}-ia_{2}& a_{3}+ia_{4}& 0 & 0 \\
	a_{3}-ia_{4}& -a_{1}-ia_{2}& 0 & 0  
	\end{array}
	\right)\,, \\
\mf{m}_{_S}= \left(
	\begin{array}{cccc}
		0 & 0 & b_{1}+ib_{2}& b_{3}+ib_{4}\\
		0 & 0 & -b_{3}+ib_{4}& b_{1}-ib_{2}\\
		-b_{1}+ib_{2}& b_{3}+ib_{4}& 0 & 0 \\
		-b_{3}+ib_{4}& -b_{1}-ib_{2}& 0 & 0  
	\end{array}
	\right)\,.
\ea
\ee
Unphysical fluctuations lie in the subspace $\mf{h}$ of $\mf{g}$ , which should be gauged away or integrated out from the fluctuation Lagrangian of the reduced theory,   
\be
\ba{c}
\mf{h}=\left( \ba{cc}
	\mf{h}_{_A} & 0 \\
	0 & \mf{h}_{_S}
\ea \right) \,, \\
\mf{h}_{_A}=\left(\ba{cccc}
i c_1 & c_2+i c_3&0&0
\\-h_c+i c_3&-i c_1&0&0
\\0&0&i c_4 & c_5+i c_6
\\0&0&-c_5+i c_6&-i c_4\ea\right)\, , \\
\mf{h}_{_S}=\left(\ba{cccc}
i d_{1} & d_2+i d_3&0&0
\\-d_2+i d_3&-i d_1&0&0
\\0&0&i d_4 & d_5+i d_6
\\0&0&-d_5+i d_6&-i d_4\ea\right)\, .
\ea
\ee
The $\kappa$-symmetry allows us to set fermionic fields to take values in $\mf{f}^{\parallel}$, 
\begin{gather}
	\mf{f}_1^{\parallel}=\left(
	\begin{array}{cc}
		0& \mf{X}_1 \\
		\mf{Y}_1& 0 
	\end{array}
	\right) \,, ~~~
	\mf{f}_3^{\parallel}=\left(
	\begin{array}{cc}
		0& \mf{X}_3  \\
		\mf{Y}_3& 0 
	\end{array}
	\right) \,,
\end{gather}
where
\begin{gather}
	\mf{X}_1=\left(
	\begin{array}{cccc}
		 0 & 0 &\alpha_1+i\alpha_2 &\alpha_3+i\alpha_4 \\
		 0 & 0 &-\alpha_3+i\alpha_4 &\alpha_1-i\alpha_2 \\
		\alpha_5+i\alpha_6 &\alpha_7-i\alpha_8 &0&0\\
		\alpha_7+i\alpha_8 &-\alpha_5+i\alpha_6 &0 &0 \\
	\end{array}
	\right)\,, \\
	\mf{Y}_1=\left(
	\begin{array}{cccc}
		0 & 0 &-\alpha_6 -i\alpha_5 & -\alpha_8 -i\alpha_7 \\
		0 & 0 &\alpha_8 -i\alpha_7 & -\alpha_6 +i\alpha_5 \\
		\alpha_2 +i\alpha_1 & \alpha_4 -i\alpha_3 &0 &0  \\
		\alpha_4 +i\alpha_3 & -\alpha_2 +i\alpha_1&0 &0  
	\end{array}
	\right)\,,
\end{gather} 
and
\small
\begin{gather}
	\mf{X}_3=\left(
	\begin{array}{cccc}
		0 & 0 &\beta_1+i\beta_2  &\beta_3+i\beta_4  \\
		0 & 0 &\beta_3-i\beta_4   &-\beta_1+i\beta_2 \\
		\beta_5+i\beta_6  &-\beta_7+i\beta_8  &0&0\\
		\beta_7+i\beta_8    &\beta_5-i\beta_6  &0 &0 
	\end{array}
	\right)\,,   \\
	\mf{Y}_3=\left(
	\begin{array}{cccc}
		0 & 0 &-\beta_6 -i\beta_5 & -\beta_8 -i\beta_7 \\
		0 & 0 &-\beta_8 +i\beta_7   & \beta_6 -i\beta_5  \\
		\beta_2 +i\beta_1 & -\beta_4 +i\beta_3 &0 &0  \\
		\beta_4 +i\beta_3 & \beta_2 -i\beta_1  &0  &0  
	\end{array}
	\right)\, . 
\end{gather}
When we derive the fluctuation Lagrangian for fermionic fluctuations, we rescale components of $\delta \Psi_{_L} \in \mf{f}_3^{\parallel}$ by $v$ or $v^*$.

\section{Pulsating string in reduced theory\label{apppuls}}

The equivalence of equations of motion for the quadratic fluctuations in the original string theory in conformal gauge and in the reduced theory was shown for an arbitrary classical solution localized in the ${\rm AdS}_2 \times {\rm S}^2$ subsector in \cite{hit}. In this sector there are many classical string configurations; the pointlike string moving along a big circle of ${\rm S}^2$, the unstable wrapping static string and inhomogeneous solutions such as pulsating strings, folded strings and magnons, e.g., \cite{hofmal, okasuz, hosv, klomac, krutse, mincir, afzfinite}. 
In this appendix we shall consider the pulsating string in $R \times {\rm S}^2$ as a nontrivial example in this subsector, and derive the Lagrangian for its quadratic fluctuations which should be directly compared with the fluctuation Lagrangian found by perturbing the Nambu action in the original string theory. 
 
In section \ref{secfold} we have shown the fluctuations from the Nambu action are related to those of the tanh model by the T-duality transformation and the same as those of the coth model. 
This variety in the reduced theory originates from the freedom in introducing the second field, i.e., $\theta_{\!_A}$ for the tanh model or $\chi_{\!_A}$ for the coth model. 
Because bosonic string theory in $R \times {\rm S}^2$ is reduced to the sin-Gordon model which possesses a single field $\phi_{\!_S}$, it is expected that the Lagrangians in the original theory and in the reduced theory exactly match.

The pulsating string solution in ${\rm S}^2$ studied in \cite{krutse, mincir} is expressed in terms of the embedding coordinates for ${\rm AdS}_5 \times {\rm S}^5$, 
\be \label{solpuls}\ba{c}
	Y_0+iY_5= e^{i\kappa \tau }\,,~~~Y_1=Y_2=0\,,\\
	X_1 + X_2= \sin \psi (\tau) e^{i m \sigma } \,, ~~~X_3=\cos \psi (\tau) \,,~~~X_4=X_5=X_6=0\,,
\ea\ee
where $m$ is the winding number on ${\rm S}^2$. 
The Virasoro constraints give one nontrivial constraint, 
\be
	\dot \psi ^2+m^2 \sin ^2\psi(\tau) =\kappa^2\,,
\ee
and the equation of motion is also derived by its derivative, 
\be
	\ddot \psi ^2+\frac{m^2}{2} \sin 2 \psi(\tau) =0\,,
\ee
where the dot represents the derivative with respect to $\tau$.

The detail of the reduction for the case of the ${\rm AdS}_2 \times {\rm S}^2$ subsector is described in \cite{hit}. 
In the deformed gWZW model the ${\rm AdS}_5$ part of the corresponding element in $G$  takes a trivial form, ${\rm diag}(i,-i,i,-i)$,  
while the ${\rm S}^5$ part is given by 
\be g=
\left(
\begin{array}{cccc}
 i \cos\phi_{\!_S} & 0 & 0 & i \sin \phi_{\!_S} \\
 0 & -i \cos\phi_{\!_S} & i \sin \phi_{\!_S} & 0 \\
 0 & i \sin \phi_{\!_S} & i \cos\phi_{\!_S} & 0 \\
 i \sin \phi_{\!_S} & 0 & 0 & -i \cos\phi_{\!_S}
\end{array}
\right) \,,
\ee
where 
\be\label{reds2}
	\partial_+ X_a \partial_- X^a =\kappa ^2 \cos 2 \phi _{\!_S} \,.
\ee
Here we used the fact the mass scale of  the reduced theory is $\mu= \kappa$. 
One can check that the classical gauge fields $A_{\pm0}$ totally vanish by solving the gauge field equation \eqref{redgaugef}. 
Substituting \eqref{solpuls} into \eqref{reds2}, then we obtain the classical solution in the ${\rm S}^5$ sector, 
  \be \label{g0puls} \ba{c}g_0 =
\left(
\begin{array}{cccc}
 \frac{i \dot\psi}{\kappa } & 0 & 0 & i \sqrt{1-\frac{\dot\psi^2}{\kappa ^2}} \\
 0 & -\frac{i \dot\psi}{\kappa } & i \sqrt{1-\frac{\dot\psi^2}{\kappa ^2}} & 0 \\
 0 & i \sqrt{1-\frac{\dot\psi^2}{\kappa ^2}} & \frac{i \dot\psi}{\kappa } & 0 \\
 i \sqrt{1-\frac{\dot\psi^2}{\kappa ^2}} & 0 & 0 & -\frac{i \dot\psi}{\kappa }
\end{array}
\right)\,, ~~~A_{\pm 0}=0\,.
\ea \ee

\subsection{Bosonic fluctuations}

Let us first discuss the bosonic fluctuations. 
Because the ${\rm AdS}_5$ part of the classical solution takes the trivial form in the present case, 
the bosonic fluctuations in the ${\rm AdS}_5$ sector are massive fields with $m_B^2= \kappa^2$.  
Then their Lagrangian is 
\be \label{pulslads}
	\mathcal{L}_1 = 2 \sum_{i=1}^4 \left( \partial_+a_i \partial_- a_i - \kappa ^2 a_i^2 \right)\,. 
\ee
Introduce the components fields of the ${\rm S}^5$ part of the physical fluctuation $\eta ^{\parallel}$ as follows,\footnote{
Both the ${\rm AdS}_5$ and ${\rm S}^5$ subsectors of $H$ are $[SU(2)]^2$. Hence component fields of the unphysical parts, $\eta ^\perp$ and $\delta A_{\pm}$, are introduced in the same way as the case of ${\rm AdS}_3  \times {\rm S}^1$ in \eqref{etao} and \eqref{dgau}. } 
\be
	\eta^\parallel= \left(
	\begin{array}{cccc}
		0 & 0 & b_{1}+ib_{2}& b_{3}+ib_{4}\\
		0 & 0 & -b_{3}+ib_{4}& b_{1}-ib_{2}\\
		-b_{1}+ib_{2}& b_{3}+ib_{4}& 0 & 0 \\
		-b_{3}+ib_{4}& -b_{1}-ib_{2}& 0 & 0  
	\end{array}
	\right)\,.
\ee
As shown in \cite{hit} one can partially fix the $H$ gauge such that the physical fields decouple from the unphysical fields. 
Then the resulting Lagrangian for the physical fluctuations is, 
\be \label{pulsls5}
\mathcal{L}_2=2 \left[ \sum _{i=1}^3\left( \partial_+b_i \partial_- b_i -\left(2 \dot\psi^2-\kappa ^2\right)b_i^2\right) + \partial_+b_4 \partial_-b_4 - \kappa^2
\left( 1-\frac{2 m^2}{\kappa ^2-\dot\psi^2}\right) b_4^2
\, \right] \,.
\ee
The sum of these two Lagrangians in the reduced theory, $\mathcal{L}_1+\mathcal{L}_2$, is exactly the same as the Lagrangian for the quadratic fluctuations found by perturbing the Nambu action in the original string theory. 

\subsection{Fermionic fluctuations}

The agreement of the bosonic fluctuations in the original theory and reduced theory requires that of fermionic fluctuations. 
So we shall next check this for completeness. 
We write down the fermionic fluctuation Lagrangian in terms of the component fields of the fermionic fluctuations introduced in \eqref{ferRL1} - \eqref{ferRL5} with $v=1$,  
\be \label{ferpuls}\ba{rl}
	{\cal L}_{\rm F} \ = \! & 2 \Big[ \sum _{i=1}^8\left( \alpha_i \dm \alpha_i +\beta_i \dm \beta_i \right) \\
	& + 2  \dot \psi  \left( -\alpha_2\beta_1  + \alpha_1\beta_2  +\alpha_4\beta_3  -\alpha_3\beta_4  +\alpha_6\beta_5  -\alpha_5\beta_6 - \alpha_8\beta_7 + \alpha_7\beta_8  \right)  \Big] \,.
\ea\ee
This Lagrangian can be separated into four parts and each of them is rewritten in terms of a real four component spinor $\Psi$, 
\be
\mathcal{L}_{\rm f}= \bar\Psi \gamma^a \partial_a\Psi  -  \dot\psi   \bar\Psi\Gamma_2\Psi \,.
\ee
So we find the fermionic fluctuations have the mass term with a coefficient $ \dot \psi $, which again agrees with the string theory result.


\begin{thebibliography}{99}

\bi{pohl}
  K.~Pohlmeyer,
  ``Integrable Hamiltonian Systems And Interactions Through Quadratic
  Constraints,''
  Commun.\ Math.\ Phys.\  {\bf 46}, 207 (1976).


\bi{devega}
  H.~J.~De Vega and N.~G.~Sanchez,
``Exact Integrability Of Strings In D-Dimensional De Sitter Space-Time,''
  Phys.\ Rev.\  D {\bf 47}, 3394 (1993);
  
F.~Combes, H.~J.~de Vega, A.~V.~Mikhailov and N.~G.~Sanchez,
  ``Multistring solutions by soliton methods in de Sitter space-time,''
  Phys.\ Rev.\  D {\bf 50}, 2754 (1994)
  [arXiv:hep-th/9310073].



\bi{hofmal}
  D.~M.~Hofman and J.~M.~Maldacena,
  ``Giant magnons,''
  J.\ Phys.\ A  {\bf 39}, 13095 (2006)
  [arXiv:hep-th/0604135].

\bi{dor}
H.~Y.~Chen, N.~Dorey and K.~Okamura,
  ``Dyonic giant magnons,''
  JHEP {\bf 0609}, 024 (2006)
  [arXiv:hep-th/0605155].

\bi{okasuz} 
  K.~Okamura and R.~Suzuki,
  ``A perspective on classical strings from complex sine-Gordon solitons,''
  Phys.\ Rev.\  D {\bf 75}, 046001 (2007)
  [arXiv:hep-th/0609026].
  
 
\bi{hosv} 
  H.~Hayashi, K.~Okamura, R.~Suzuki and B.~Vicedo,
  ``Large Winding Sector of AdS/CFT,''
  JHEP {\bf 0711}, 033 (2007)
  [arXiv:0709.4033].
  
  
  
\bi{klomac}
  T.~Klose and T.~McLoughlin,
  ``Interacting finite-size magnons,''
  J.\ Phys.\ A  {\bf 41}, 285401 (2008)
  [arXiv:0803.2324].
  
    \bi{jevjin1}
A.~Jevicki, K.~Jin, C.~Kalousios and A.~Volovich,
  ``Generating AdS String Solutions,''
  JHEP {\bf 0803}, 032 (2008)
  [arXiv:0712.1193];
  
A.~Jevicki and K.~Jin,
  ``Solitons and AdS String Solutions,''
  Int.\ J.\ Mod.\ Phys.\  A {\bf 23}, 2289 (2008)
  [arXiv:0804.0412].

 \bi{jevjin2}
  I.~Aniceto and A.~Jevicki,
  ``N-body Dynamics of Giant Magnons in ${\rm R} \x {\rm S}^2$,''
  arXiv:0810.4548;
  
  A.~Jevicki and K.~Jin,
  ``Moduli Dynamics of ${\rm AdS}_3$ Strings,''
  JHEP {\bf 0906}, 064 (2009)
  [arXiv:0903.3389].



\bi{mira}
J.~L.~Miramontes,
  ``Pohlmeyer reduction revisited,''
  JHEP {\bf 0810}, 087 (2008)
  [arXiv:0808.3365].
  
T.~J.~Hollowood and J.~L.~Miramontes,
  ``Magnons, their Solitonic Avatars and the Pohlmeyer Reduction,''
  JHEP {\bf 0904}, 060 (2009)
  [arXiv:0902.2405].
  
  T.~J.~Hollowood and J.~L.~Miramontes,
  ``A New and Elementary ${\rm CP}^n$ Dyonic Magnon,''
  JHEP {\bf 0908}, 109 (2009)
  [arXiv:0905.2534].





\bi{almal1}
  L.~F.~Alday and J.~Maldacena,
  ``Minimal surfaces in AdS and the eight-gluon scattering amplitude at strong
  coupling,''
  arXiv:0903.4707;

  ``Null polygonal Wilson loops and minimal surfaces in Anti-de-Sitter space,''
  JHEP {\bf 0911}, 082 (2009)
  [arXiv:0904.0663].
  







\bi{almal2}

  L.~F.~Alday, D.~Gaiotto and J.~Maldacena,
  ``Thermodynamic Bubble Ansatz,''
  arXiv:0911.4708.


    L.~F.~Alday, J.~Maldacena, A.~Sever and P.~Vieira,
  ``Y-system for Scattering Amplitudes,''
  arXiv:1002.2459.

  \bi{misu1}
  H.~Dorn,
  ``Some comments on spacelike minimal surfaces with null polygonal boundaries
  in ${\rm AdS}_m$,''
  JHEP {\bf 1002}, 013 (2010)
  [arXiv:0910.0934].
  
  H.~Dorn, N.~Drukker, G.~Jorjadze and C.~Kalousios,
  ``Space-like minimal surfaces in ${\rm AdS} \x {\rm S}$,''
  JHEP {\bf 1004}, 004 (2010)
  [arXiv:0912.3829].


\bi{misu2}
  A.~Jevicki and K.~Jin,
  ``Series Solution and Minimal Surfaces in AdS,''
  JHEP {\bf 1003}, 028 (2010)
  [arXiv:0911.1107].

  

\bi{misu3}
  B.~A.~Burrington and P.~Gao,
  ``Minimal surfaces in AdS space and Integrable systems,''
  JHEP {\bf 1004}, 060 (2010)
  [arXiv:0911.4551].








\bi{gt1}
  M.~Grigoriev and A.~A.~Tseytlin,
  ``Pohlmeyer reduction of ${\rm AdS}_5 \x {\rm S}^5$ superstring sigma model,''
  Nucl.\ Phys.\  B {\bf 800} (2008) 450
  [arXiv:0711.0155].
\bi{mikhshaf}
  A.~Mikhailov and S.~Schafer-Nameki,
  ``Sine-Gordon-like action for the Superstring in ${\rm AdS}_5 \x {\rm S}^5$,''
  JHEP {\bf 0805}, 075 (2008)
  [arXiv:0711.0195].
\bi{gt2}  M.~Grigoriev and A.~A.~Tseytlin,
  ``On reduced models for superstrings on ${\rm AdS}_n \x {\rm S}^n$,''
  Int.\ J.\ Mod.\ Phys.\  A {\bf 23} (2008) 2107
  [arXiv:0806.2623].


\bi{mt}
R.~R.~Metsaev and A.~A.~Tseytlin,
  ``Type IIB superstring action in ${\rm AdS}_5 \x {\rm S}^5$ background,''
  Nucl.\ Phys.\  B {\bf 533}, 109 (1998)
  [arXiv:hep-th/9805028].



   \bi{rtt}
  R.~Roiban, A.~Tirziu and A.~A.~Tseytlin,
  ``Two-loop world-sheet corrections in AdS5 x S5 superstring,''
  JHEP {\bf 0707}, 056 (2007)
  [arXiv:0704.3638].

 
  
  
\bi{rotecusp}
R.~Roiban and A.~A.~Tseytlin,
  ``Strong-coupling expansion of cusp anomaly from quantum superstring,''
  JHEP {\bf 0711}, 016 (2007)
  [arXiv:0709.0681].
  
  \bi{rt2}
  R.~Roiban and A.~A.~Tseytlin,
  ``Spinning superstrings at two loops: strong-coupling corrections to dimensions of large-twist SYM operators,''
  Phys.\ Rev.\  D {\bf 77} (2008) 066006
  [arXiv:0712.2479].




\bi{rtfin}
  R.~Roiban and A.~A.~Tseytlin,
  ``UV finiteness of Pohlmeyer-reduced form of the ${\rm AdS}_5 \x {\rm S}^5$ superstring theory,''
  JHEP {\bf 0904} (2009) 078
  [arXiv:0902.2489].








  \bi{hit}
   B.~Hoare, Y,~Iwashita and A.~A.~Tseytlin,
  ``Pohlmeyer-reduced form of string theory in ${\rm AdS}_5 \x {\rm S}^5$: Semiclassical expansion,''
 J.\ Phys.\ A {\bf 42} (2009) 375204
  [arXiv:0906.3800].

\bi{gkp}
  S.~S.~Gubser, I.~R.~Klebanov and A.~M.~Polyakov,
  ``A semi-classical limit of the gauge/string correspondence,''
  Nucl.\ Phys.\  B {\bf 636}, 99 (2002)
  [arXiv:hep-th/0204051].








 
  \bi{ft1}
   S.~Frolov and A.~A.~Tseytlin,
  ``Semiclassical quantization of rotating superstring in ${\rm AdS}_5 \x {\rm S}^5$,''
  JHEP {\bf 0206},  (2002) 007
  [arXiv:hep-th/0204226].
  
  \bi{ftt}
  S.~Frolov, A.~Tirziu and A.~A.~Tseytlin,
  ``Logarithmic corrections to higher twist scaling at strong coupling from AdS/CFT,''
  Nucl.\ Phys.\  B {\bf 766} (2007) 232
  [arXiv:hep-th/0611269].
  
  
  \bibitem{bdfpt}
  M.~Beccaria, G.~V.~Dunne, V.~Forini, M.~Pawellek and A.~A.~Tseytlin,
  ``Exact computation of one-loop correction to energy of spinning folded string in ${\rm AdS}_5 \x {\rm S}^5$,''
  J.\ Phys.\ A {\bf 43}, 165402 (2010)
  [arXiv:1001.4018].


  



\bi{art}
G. Arutyunov, J. Russo and A. A. Tseytlin
``Spinning strings in ${\rm AdS}_5 \x {\rm S}^5$: New integrable system relations,''
Phys.\ Rev.\ D {\bf 69} (2004) 086009 
 [arXiv:hep-th/0311004].

 
\bi{ptt}
  I.~Y.~Park, A.~Tirziu and A.~A.~Tseytlin,
  ``Spinning strings in ${\rm AdS}_5 \x {\rm S}^5$: One-loop correction to energy in  SL(2)
  sector,''
  JHEP {\bf 0503}, 013 (2005)
  [arXiv:hep-th/0501203].



\bi{gv}
N. Gromov and P. Vieira
``The ${\rm AdS}_5 \x {\rm S}^5$ superstring quantum spectrum from the algebraic curve,''
Nucl.\ Phys.\ B {\bf 789} (2008) 175-208
 [arXiv: hep-th/0703191]
 
 
\bi{mikhcir}
   V.~Mikhaylov,
  ``On the Fermionic Frequencies of Circular Strings,''
  arXiv:1002.1831.

 
 
 \bi{krtt}
M.~Kruczenski, R.~Roiban, A. Tirziu and A.~A.~Tseytlin,
``Strong-coupling expansion of cusp anomaly and gluon amplitudes from quantum open strings in ${\rm AdS}_5 \x {\rm S}^5$,''
Nucl. Phys. B {\bf 791}, 93 (2008)
 [arXiv:0707.4254]. 



\bi{kru}
M.~Kruczenski,
``A Note on twist two operators in N=4 SYM and Wilson loops in Minkowski signature,''
JHEP {\bf 0212}, (2002) 024 
 [arXiv:hep-th/0210115].

 
 \bi{krusp}
 M.~Kruczenski,
  ``Spiky strings and single trace operators in gauge theories,''
  JHEP {\bf 0508}, 014 (2005)
  [arXiv:hep-th/0410226].
  
  \bi{iktt}
    R.~Ishizeki, M.~Kruczenski, A.~Tirziu and A.~A.~Tseytlin,
  ``Spiky strings in ${\rm AdS}_3 \x {\rm S}^1$ and their AdS-pp-wave limits,''
  Phys.\ Rev.\  D {\bf 79}, 026006 (2009)
  [arXiv:0812.2431].
  
  

  
  \bibitem{grrtv}
  S.~Giombi, R.~Ricci, R.~Roiban, A.~A.~Tseytlin and C.~Vergu,
  ``Generalized scaling function from light-cone gauge ${\rm AdS}_5 \x {\rm S}^5$
  superstring,''
  arXiv:1002.0018.



\bi{krutse}
  M.~Kruczenski and A.~A.~Tseytlin,
  ``Semiclassical relativistic strings in $ {\rm S}^5$ and long coherent operators in N = 4 SYM theory,''
  JHEP {\bf 0409}, 038 (2004)
  [arXiv:hep-th/0406189].









\bi{mincir}
  J.~A.~Minahan,
  ``Circular semiclassical string solutions on ${\rm AdS}_n \x {\rm S}^n$,''
  Nucl.\ Phys.\  B {\bf 648}, 203 (2003)
  [arXiv:hep-th/0209047].


\bi{krutir} 
  M.~Kruczenski and A.~Tirziu,
  ``Spiky strings in Bethe Ansatz at strong coupling,''
  Phys.\ Rev.\  D {\bf 81}, 106004 (2010)
  [arXiv:1002.4843].



\bi{afzfinite}
  G.~Arutyunov, S.~Frolov and M.~Zamaklar,
  ``Finite-size effects from giant magnons,''
  Nucl.\ Phys.\  B {\bf 778}, 1 (2007)
  [arXiv:hep-th/0606126].


 
\end{thebibliography}
\end{document}